\documentclass[aps,english,pra,floatfix,amsmath,superscriptaddress,tightenlines,twocolumn,nofootinbib,accepted=2023-07-18,a4paper]{quantumarticle}
\pdfoutput=1
\usepackage{mathtools,amsthm,amssymb,setspace}
\usepackage[pass]{geometry}
\usepackage{mathrsfs}%
\usepackage{stmaryrd}
\usepackage{graphicx}
\usepackage{tabularx}%
\usepackage{hyperref}
\hypersetup{colorlinks=true,linkcolor=red,citecolor=magenta,urlcolor=blue}
\usepackage{lipsum}    
\usepackage{tikz-cd}
\usepackage{tikz}
\usepackage{soul}
\usepackage{tikz-3dplot}
\usetikzlibrary{quantikz}
\usetikzlibrary{spy}
\usetikzlibrary{arrows.meta}
\usetikzlibrary{calc}
\usetikzlibrary{decorations.pathreplacing,calligraphy}
\usepackage{intcalc}

\usepackage[shortlabels]{enumitem}
\usepackage{soul}
\usepackage[utf8]{inputenc}
\usepackage{xcolor}
\usepackage{float}
\usepackage[newcommands]{ragged2e}
\usepackage{comment}

\usepackage{caption}
\usepackage{subcaption}
\newcolumntype{b}{X}
\newcolumntype{s}{>{\hsize=.5\hsize}X}

\makeatletter
\renewcommand\listoffigures{\@starttoc{lof}}
\renewcommand\listoftables{\@starttoc{lot}}
\renewcommand{\tablename}{Table}
\renewcommand{\figurename}{Figure}
\makeatother

\DeclarePairedDelimiter\floor{\lfloor}{\rfloor}
\newcommand{\be}{\begin{equation}}
\newcommand{\ee}{\end{equation}}
\definecolor{pink}{rgb}{0.72,0.18,0.95}
\definecolor{celeste}{rgb}{0.09, 0.81, 0.9}
\definecolor{emerald}{rgb}{0, 0.677778, 0.555556}
\newtheorem{theorem}{Theorem}[section]
\theoremstyle{definition}
\newtheorem{definition}[theorem]{Definition}

\begin{document}
\date{2023-08-05}
\title{Concatenation Schemes for Topological Fault-tolerant Quantum Error Correction}

 \author{Zhaoyi Li}
 \email{ladmon@stanford.edu}
 \affiliation{Department of Physics, Stanford University, Stanford, CA 94305, USA}
  
\author{Isaac Kim}
\email{ikekim@ucdavis.edu}
\affiliation{Department of Computer Science, University of California, Davis, CA 95616, USA}
 
\author{Patrick Hayden}
\email{phayden@stanford.edu}
\affiliation{Department of Physics, Stanford University, Stanford, CA 94305, USA}

\begin{abstract}
We investigate a family of fault-tolerant quantum error correction schemes based on the concatenation of small error detection or error correction codes with the three-dimensional cluster state. We propose fault-tolerant state preparation and decoding schemes that effectively convert every circuit-level error into an erasure error, leveraging the cluster state's high threshold against such errors. We find a set of codes for which such a conversion is possible, and study its performance against the standard circuit-level depolarizing model. Our best performing scheme, which is based on a concatenation with a classical code, improves the threshold by $16.5\%$ and decreases the spacetime overhead by $32\%$ compared to the scheme without concatenation, with each scheme subject to a physical error rate of $10^{-3}$ and achieving a logical error rate of $10^{-6}$.

\end{abstract}

\maketitle

\section{Introduction}
\label{sec:intro}

Quantum computers are expected to provide substantial speedups for a number of practical problems of interest, such as factoring, quantum chemistry, and the simulation of physical models that describe nature~\cite{cao2019quantum,shor1994algorithms,georgescu2014quantum}. However, the ubiquitous presence of noise makes qubits very fragile and easily affected by their environment, which can cause them to decohere.

The fault tolerance threshold theorem~\cite{aliferis2005quantum,dennis2002topological,bravyi2020quantum,spedalieri2008latency,cross2007comparative} provides a foundation
for the long-term development of quantum computers despite the presence of noise. This theorem states that there is a finite error rate below which a noisy quantum computer can simulate a noiseless quantum computer with an error rate that can be made to be arbitrarily small, while only paying a polylogarithmic overhead in space and time. More practically, this theorem asserts that building an effectively noiseless quantum computer is in principle possible.

However, in order to build such a quantum computer, one must develop a fault-tolerant quantum computation scheme that is experimentally achievable. The figures of merit that are often considered in the literature can be divided into three categories: threshold, overhead, and connectivity. The threshold is an error rate below which scaling up the system decreases the logical error rate. As such, it is desirable to have a system with a higher threshold. Overhead is the ratio of the number of physical qubits to logical qubits, which is a metric best minimized if possible. Lastly, connectivity refers to the set of gates that can be applied between the qubits in the quantum computer. For this last category, it is desirable to have requirements compatible with the capabilities of the actual hardware. 

Kitaev's surface code~\cite{dennis2002topological} has been the major focus of research on practical fault tolerance over the past two decades. The surface code boasts a high threshold and can be realized with qubits connected in a two-dimensional square lattice, making it the leading candidate architecture for early-generation fault-tolerant quantum computers. 

A more traditional approach to fault tolerance was based on concatenated codes. While the associated codes form the major underpinning of the early theory of fault tolerance, the estimated threshold of the schemes tends to be lower; see Ref.~\cite{chamberland2016thresholds} for example. However, one notable exception to this rule is Knill's fault tolerance scheme, which is based on postselection~\cite{knill2004fault,knill2004fault1,knill2005quantum}.

Postselection is a technique in which one accepts an outcome only if a certain set of criteria is satisfied. For example, in magic state distillation, one accepts the distilled magic state only if all the syndrome measurements report $+1$. Knill's key insight was that the threshold for postselection schemes tends to be higher than fault-tolerant quantum error correction schemes. Of course, if one relies on postselection alone, the probability of accepting the result of a post-selected computation decreases exponentially with the number of post-selected operations, making this approach impractical~\cite{knill2004fault,knill2004fault1}. However, what one can do is concatenate such a scheme with another code that can deal with the events in which the postselection procedure does not accept the outcome. 

While Knill's scheme is based on the concatenation of a postselection scheme with a fault-tolerant concatenated quantum error correction scheme, there is a viable alternative: concatenating a postselection scheme with the surface code, or equivalently the three-dimensional (3D) cluster state. It is well-known that these schemes~\cite{stace2011loss} have high thresholds against loss compared to Pauli noise. Therefore, one can hope to improve the logical error rate by converting weight-$1$ Pauli errors to erasure errors, by declaring loss if the postselection scheme does not accept. The main purpose of this paper is to provide a careful study of this idea. 

This paper reports a comprehensive study in this direction for a variety of small quantum error correcting codes concatenated with the 3D cluster state~\cite{raussendorf2006fault,raussendorf2007fault,raussendorf2007topological}. The codes we employ can be broadly classified into two categories. The first category, which we refer to as Type I codes, involves codes for which the concatenation can be implemented straightforwardly using the fault-tolerant logical gates available to those codes. While we observe a general increase in threshold against the phenomenological noise model, these codes are less effective against circuit-level noise. The second category, which we refer to as Type II codes, are codes for which some of the logic gates used are not obviously fault-tolerant in the sense that a single error can propagate to multiple errors in the other code block. Somewhat surprisingly, some of these codes outperform the bcc cluster state, even against circuit-level noise.

We attribute this fact to a surprising property of our state preparation circuit: any single-qubit circuit-level noise operator as well as the two-qubit circuit-level noise operator that are implemented with the CZ gates propagate, up to stabilizers, to a tensor product of errors on disjoint code blocks which are detectable from the syndrome information within that block. In particular, upon measuring this code block, one can convert the Pauli error into an erasure error. Importantly, via this process, \emph{every} single-qubit circuit-level noise operator propagates to an erasure error of the 3D cluster state. Because the logical gates used for the codes of Type II are not obviously fault-tolerant, the fact that we can achieve this conversion is not obvious. 

We note that there has been significant recent work studying a concatenation of the 3D cluster state, or equivalently the 2D surface code, with more hardware-efficient codes~\cite{wang2019quantum, bourassa2021blueprint,noh2022low,fukui2018high,yamasaki2020polylog,noh2020fault,vuillot2019quantum,wu2022erasure,kubica2022erasure}. 
While the underlying idea is similar, there is a crucial difference. The prior work exploits some special structures in the noise or concatenates with a more hardware-specific scheme, e.g., the Gottesman-Kitaev-Preskill (GKP) code. On the other hand, our scheme does not make these extra assumptions. In particular, it should be possible to concatenate these schemes with our scheme, which may lead to further improvements.

The rest of this paper is structured as follows. In Section \ref{sec:background}, we introduce stabilizer codes, cluster states, and concatenation schemes. In Section \ref{sec:concatenation}, we demonstrate two types of concatenation schemes for fault-tolerant quantum error correction. In Section \ref{sec:sim}, numerical simulations are performed to study the threshold error rate of these schemes. Finally, in Section \ref{sec:analytics}, we use numerical and analytical results to explain the low-error behavior of the qubit overhead.

\section{Background}
\label{sec:background}
In this section, we review known facts about quantum error correction that are relevant to this work.
\subsection{Stabilizer codes}
\label{subsec:stabilizer}
A \emph{stabilizer code} is a subspace spanned by the joint $+1$ eigenstates of the elements of the \emph{stabilizer group}~\cite{Gottesman1997}. A stabilizer group is a commutative subgroup of the Pauli group which does not contain $-I^{\otimes n}$ as an element. Following the usual convention, we shall say a code is an $\llbracket n,k,d\rrbracket$-code if the stabilizer group is over $n$ qubits, the number of logical qubits is $k$, and the code distance is $d$.

Stabilizer codes are a flexible, ubiquitous foundation upon which one can build schemes for fault-tolerant quantum computation~\cite{Aharonov2008,gottesman1998theory}. There are several approaches to achieve that aim, using block codes~\cite{gottesman1998theory,Steane1999,Steane2005,Gottesman2014}, topological codes~\cite{dennis2002topological,raussendorf2006fault,raussendorf2007topological,raussendorf2007fault}, and code concatenation~\cite{Aharonov2008,aliferis2005quantum,knill2005quantum}. In this paper, we will investigate a hybrid of the latter two. 

The topological code of particular relevance to this paper is Kitaev's surface code~\cite{Kitaev2003}. This code boasts a high threshold and planar connectivity, making it one of the leading approaches to building a large-scale fault-tolerant quantum computer. It is well-known that there is an intimate relationship between the surface code and the 3D cluster state, which is a resource state for fault-tolerant quantum computation~\cite{raussendorf2006fault,raussendorf2007topological,raussendorf2007fault}. 

A \emph{cluster state} is a quantum state defined in terms of a graph $G=(V,E)$, where $V$ is the set of vertices and $E$ is a set of edges. More specifically, the cluster state $|\psi_G\rangle$ is defined as the unique state over $|V|$ qubits -- labeled in terms of $v\in V$ -- determined by the following equation:
\begin{equation}
    X_v \left( \bigotimes_{u: (v,u) \in E} Z_u \right) |\psi_G\rangle = |\psi_G\rangle,
\end{equation}
for all $v\in V$. Alternatively, $|\psi_G\rangle$ can be defined as:
\begin{equation}
    |\psi_G\rangle = \prod_{(v,u) \in E} CZ_{u,v} \left(\bigotimes_{v\in V} |+\rangle_v\right),
\end{equation}
where $|+\rangle = \frac{|0\rangle + |1\rangle}{\sqrt{2}}$ and $CZ_{u,v}$ is the controlled-$Z$ gate between $u$ and $v$. Thus, the cluster state can be created by preparing all the qubits in the $|+\rangle$ state and applying a sequence of $CZ$ gates.

\begin{figure}[h]
\begin{center}
\includegraphics[scale=0.03]{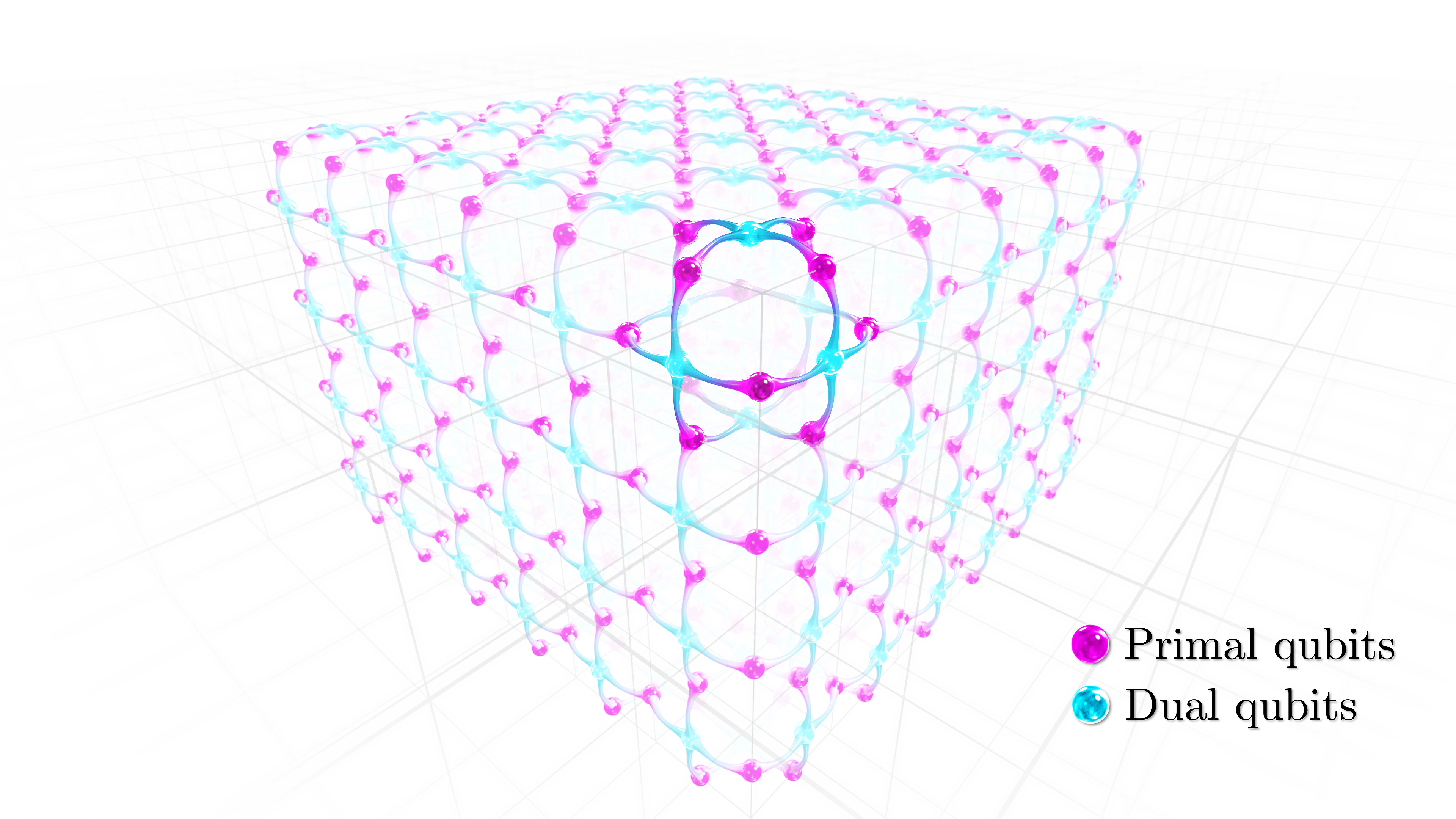}
\end{center}
\caption{Illustration of the Raussendorf lattice. There are two types of qubits in the lattice, the magenta balls represent primal qubits and the cyan balls represent dual qubits. The linkage between the balls represents $CZ$ gate applications.}
\label{fig:lattice}
\end{figure}

The 3D cluster state~\cite{Raussendorf2005,raussendorf2007topological} is a cluster state defined over a graph $G_{\text{bcc}}$, a bcc lattice; see Figure~\ref{fig:lattice}. The importance of the 3D cluster state comes from the fact that it is a resource state for fault-tolerant measurement-based quantum computation. For instance, by measuring all but the bottom and top layers of Figure~\ref{fig:lattice}, up to a Pauli correction, we can establish a logical Bell pair between two surface code states at the bottom and top layers.

It is possible to obtain the surface code from the 3D cluster state through the process of foliation~\cite{bolt2016foliated}. Moreover, the surface code state can be fault-tolerantly teleported by measuring consecutive layers of the 3D cluster state in the $X$-basis~\cite{Raussendorf2005}. In a fault-tolerant cluster state quantum computation, $Z$-basis measurements are only used for removing the vertices of the graph that defines the cluster state~\cite{briegel2009measurement}. $Y$-basis measurements are used only when applying a logical $S$-gate.~\cite{Brown_2020}. Therefore, for the purpose of storing a quantum information fault-tolerantly, it suffices to use only $X$-basis measurements.

Thus, while we shall primarily work with the 3D cluster state in this paper, the conclusions made, for instance, on the threshold and the overhead, are easily translatable to that of the surface code. We note in passing that the concatenation scheme may be applicable to other cluster states using crystalline structures~\cite{nickerson2013topological,newman2020generating}. We leave such studies for future work.

\subsection{Concatenation}
\label{subsec:concatenation_background}

In this paper, we propose a fault-tolerant quantum error correction scheme based on concatenating the 3D cluster state with a quantum error-correcting code. We aim to construct a state we denote as $|\psi_{G,\mathcal{C}}\rangle$, where $G=(V,E)$ is a graph and $\mathcal{C}$ is a stabilizer code that encodes a single logical qubit. Doing so requires performing logical operators on the logical qubit, which we will refer to as a {\it code block} of physical qubits. Specifically, given $\mathcal{C}$, let $|\overline{+}\rangle$ be its logical $|+\rangle$ state and let $\overline{CZ}_{u, v}$ be a logical $\overline{CZ}$ gate between two code blocks, labeled by $u, v\in V$. The state $|\psi_{G, \mathcal{C}}\rangle$ is defined as 
\begin{equation}
    |\psi_{G,\mathcal{C}}\rangle = \prod_{(u,v)\in E} \overline{CZ}_{u,v}\left(\bigotimes_{v\in V} |\overline{+}\rangle_v \right). \label{eq:cluster_concatenated}
\end{equation}
One can formally view $|\psi_{G,\mathcal{C}}\rangle$ as a state living in a code $\bigotimes_{v\in V}\mathcal{C}_v$.

While our main focus lies in fault-tolerant quantum error correction, let us remark that, insofar as the state in Eq.~\eqref{eq:cluster_concatenated} can be prepared, the basic ingredients of fault-tolerant quantum computation delineated in Ref.~\cite{raussendorf2007topological} can be implemented in a straightforward way. The Clifford gates could be performed with logical $\overline{X}$ and $\overline{Z}$ measurements, assisted by an ancillary $+1$ eigenstate of the Pauli-$Y$ operator~\cite{dennis2002topological}. The faulty magic state can be encoded into a code block by performing an (unprotected) magic-basis measurement~\cite{Litinski2019}, followed by distillation~\cite{Bravyi2005}. Thus, a universal set of fault-tolerant quantum gates can be implemented using the state in Eq.~\eqref{eq:cluster_concatenated}.

\section{Concatenation schemes}
\label{sec:concatenation}

In this section, we describe the concatenation schemes in detail. To explain our principle for choosing codes, we first need to explain our approach to decoding. For fault-tolerant error correction, it suffices to perform logical $X$-basis measurements for each code block, i.e. the {\it inner} codes, feed the measurement outcomes to higher-level decoding algorithms, such as the minimum-weight perfect matching (MWPM) decoder~\cite{dennis2002topological,raussendorf2007fault} or the union-find decoder~\cite{Delfosse2021}, and apply the appropriate Pauli correction. We shall refer to these algorithms as {\it outer} decoding algorithms. The standard inputs to these decoding algorithms are the $X$-basis measurement results of the individual physical qubits. Applied to our setup, one way to decode our scheme would be to perform a destructive logical $X$-basis measurement for every vertex and feed the logical measurement outcome to an outer decoding algorithm. 

However, at least when it comes to the threshold, we consider it unlikely for this approach to be able to outperform the body-centered cubic (bcc) cluster state (without concatenation). The main reason is that, to the best of our knowledge, there is no quantum error correcting code whose pseudothreshold\footnote{A pseudothreshold of a code is defined as the error rate below which the logical error rate is below the physical error rate.} is above the fault-tolerant quantum error correction threshold of the 3D cluster state. If the pseudothreshold is lower than the threshold of the bcc cluster state, the overall threshold of the scheme will be limited by it. This argument suggests that, unless more information is passed from the lower level to the outer decoder~\cite{Poulin_2006}, improvements in threshold are likely unachievable.

This is why we seek to consider a different decoding algorithm in which information from the code $\mathcal{C}$ is passed to the outer decoding algorithm. To that end, we remark on a few facts that are relevant to our choice of inner codes. First of all, both MWPM and union-find decoders can also easily accommodate erasure errors. In particular, these decoders boast a much higher threshold for erasure error than for Pauli noise. Secondly, for the bcc cluster state with the phenomenological noise model, i.e., a model in which a noiseless cluster state is subjected to an identical independent noise model, the threshold for the depolarizing noise is $\sim 3\%$ whereas the threshold against loss is $25\%$~\cite{barrett2010fault}. Thirdly, when an error is detected at a low level, instead of correcting the error, one can instead opt to artificially ``lose'' that logical qubit by erasing it. This information can be fed into the outer decoder as an erasure error. This Pauli-to-loss conversion of the error model is the main idea behind our decoding scheme.

To make this idea work, we must choose the code $\mathcal{C}$ such that there is a circuit that prepares the state $|\psi_{G_{\text{bcc}},\mathcal{C}}\rangle$ in such a way that any single circuit-level error propagates to an error convertible to a (possibly correlated) erasure error upon applying the destructive logical $X$-basis measurement of $\mathcal{C}.$ If $\mathcal{C}$ is a Calderbank-Shor-Steane (CSS) code, the $X$-basis measurement is unaffected by any $X$-error. Thus, in order to make a circuit-level error detectable from the logical $X$ measurement of $\mathcal{C}$, it suffices for the propagated error to be equivalent up to stabilizers and Pauli-$X$ operators to a tensor product of Paulis over different code blocks, such that the individual errors within each code block are detectable. More formally, we propose the following definition.

Let us first define the \emph{circuit-level error.} A circuit-level error is an error occurring after individual gates, specifically, on a set of qubits the gate acts on. For instance, if we have a noisy $CZ$ gate, its circuit-level noise will be one or two-qubit Paulis occurring at the qubits that the $CZ$ gate acts on. We now define what it means for these errors to be $\mathcal{C}$-detectable.

\begin{definition}
Let $\mathcal{C}$ be a CSS code. Consider a circuit that prepares $|\psi_{G,\mathcal{C}}\rangle$. A circuit-level noise operator is $\mathcal{C}$-detectable if the noise, up to some stabilizer and Pauli-$X$ operators $S$, is of the following form:
\begin{equation}
    S \prod_{v\in G} P_v,
\end{equation}
where $P_v$ is an error detectable with respect to $\mathcal{C}$.
\end{definition}
\begin{definition}
Let $\mathcal{C}$ be a CSS code. Consider a circuit that prepares $|\psi_{G,\mathcal{C}}\rangle$. This circuit is $\mathcal{C}$-detectable if every circuit-level Pauli noise operator is $\mathcal{C}$-detectable.
\end{definition}

There's a subtle difference between $\mathcal{C}$-detectability and conventional fault-tolerant circuits. For a fault-tolerant gadget~\cite{aliferis2005quantum}, the output must contain at most one fault, whereas $\mathcal{C}$-detectability only requires the number of faults to be within the detection ability of the inner code up to stabilizers and $X$ operators. If we can create a $\mathcal{C}$-detectable state preparation circuit, we can convert circuit-level errors to erasure errors of the outer code.

The main question we address in this Section is whether there are circuits preparing the state $|\psi_{\text{bcc},\mathcal{C}}\rangle$ in a $\mathcal{C}$-detectable way. We discuss two families of such codes. The first family consists of codes that are equipped with a fault-tolerant logical $|+\rangle$-state preparation and logical (transversal) $CZ$ gate. Such codes, by using these operations, come with a natural $\mathcal{C}$-detectable state preparation circuit. We refer to the inner codes used in these constructions as {\it Type I} codes. 

Surprisingly, we find that there are choices of inner codes without any fault-tolerant logical $\overline{CZ}$ gates, which nevertheless admit a $\mathcal{C}$-detectable state preparation circuits for $|\psi_{\text{bcc}, \mathcal{C}}\rangle$. We refer to these inner codes as {\it Type II} codes.

\subsection{Type I}
\label{sec:typeA}
We consider two examples of Type I codes: a $\llbracket4,1,1,2\rrbracket$ subsystem code~\cite{napp2012optimal} \footnote{In this case, the $\llbracket4,1,1,2\rrbracket$ code is derived from a $\llbracket4,2,2\rrbracket$ code, the $n=2$ Bacon-Shor code, by employing just one of its logical qubits.} and the Steane code~\cite{Steane1999}. The stabilizers of the $\llbracket4,1,1,2\rrbracket$ are given by $X_1X_2$, $Z_1Z_3$, $XXXX$ and $ZZZZ$, and the logical operations are given by
    \begin{equation}
        \begin{aligned}
        \overline{X} &= X_1X_3\\
        \overline{Z} &= Z_1Z_2.
        \end{aligned}
    \end{equation}
Transversal logical $\overline{CZ}$ gate construction is trivially possible. Moreover, $\llbracket4,1,1,2\rrbracket$ concatenation with the surface code, which is unitary-equivalent to a $4.8.8$ color code, has been studied previously~\cite{criger2016noise}.

Another code we consider is the $\llbracket7,1,3\rrbracket$ Steane code, whose stabilizers are $X_1X_2X_3X_4$, $X_1X_3X_5X_7$, $X_3X_4X_6X_7$, $Z_1Z_2Z_3Z_4$, $Z_1Z_3Z_5Z_7$,  and $Z_3Z_4Z_6Z_7$. This is also the smallest 2D color code~\cite{bermudez2017assessing}.

Since both codes admit transversal $CZ$ gates, $Z$ errors are local throughout the whole circuit. On the other hand, although a $CZ$ gate will propagate $X$ errors to $Z$ errors, transversality guarantees at most one error will present in a given code block, which could be effectively corrected or detected by the inner code.

\subsection{Type II}
\label{sec:typeB}
\begin{figure}[h]
    \centering
    \begin{tikzpicture}[baseline={([yshift=-.5ex]current bounding box.center)}]
    \draw[pink!75, fill=pink!40, rounded corners=2,thick]
     (-0.25, -0.25) rectangle (0.25,1.25) {};
    \draw[celeste, fill=celeste!40, rounded corners=2,thick]
     (1.75, -0.25) rectangle (2.25,1.25) {};
    \node[circle, fill=pink!75, inner sep=0pt, minimum size=0.2cm] (v1) at (0,0) {};
    \node[circle, fill=pink!75, inner sep=0pt, minimum size=0.2cm] (v2) at (0,1) {};
    \node[circle, fill=celeste, inner sep=0pt, minimum size=0.2cm] (u1) at (2,0) {};
    \node[circle, fill=celeste, inner sep=0pt, minimum size=0.2cm] (u2) at (2,1) {};
    \node[below] at (0, -0.25) {\footnotesize $\llbracket2,1,1\rrbracket$};
    \node[below] at (2, -0.25) {\footnotesize $\llbracket2,1,1\rrbracket$};

    \draw[gray] (v1) -- (u1);
    \draw[gray] (v2) -- (u2);
    \draw[gray] (v1) -- (u2);
    \draw[gray] (v2) -- (u1);
    \end{tikzpicture}
    \caption{Logical $\overline{CZ}$ gate of the $\llbracket2,1,1\rrbracket$ code. Here, two grouped dots represent the two qubits in each $\llbracket2,1,1\rrbracket$ code.}
    \label{fig:211}
\end{figure}
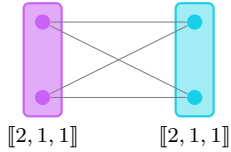

In this section, we introduce Type II codes using up to $4$ qubits. We first introduce the $\llbracket2,1,1\rrbracket$ repetition code. The stabilizer group of the $\llbracket2,1,1\rrbracket$ code is $\langle\{ X_1X_2\}\rangle$ and the logical operators are 
\begin{equation}
    \begin{aligned}
    \overline{Z} &= Z_1Z_2 \\
    \overline{X} &= X_1.
    \end{aligned}
\end{equation}
The logical $\overline{CZ}$ gate can be applied by applying a physical $CZ$ gate to every pair of connected qubits in Figure~\ref{fig:211}.

\begin{figure}
\centering\includegraphics[scale=0.033]{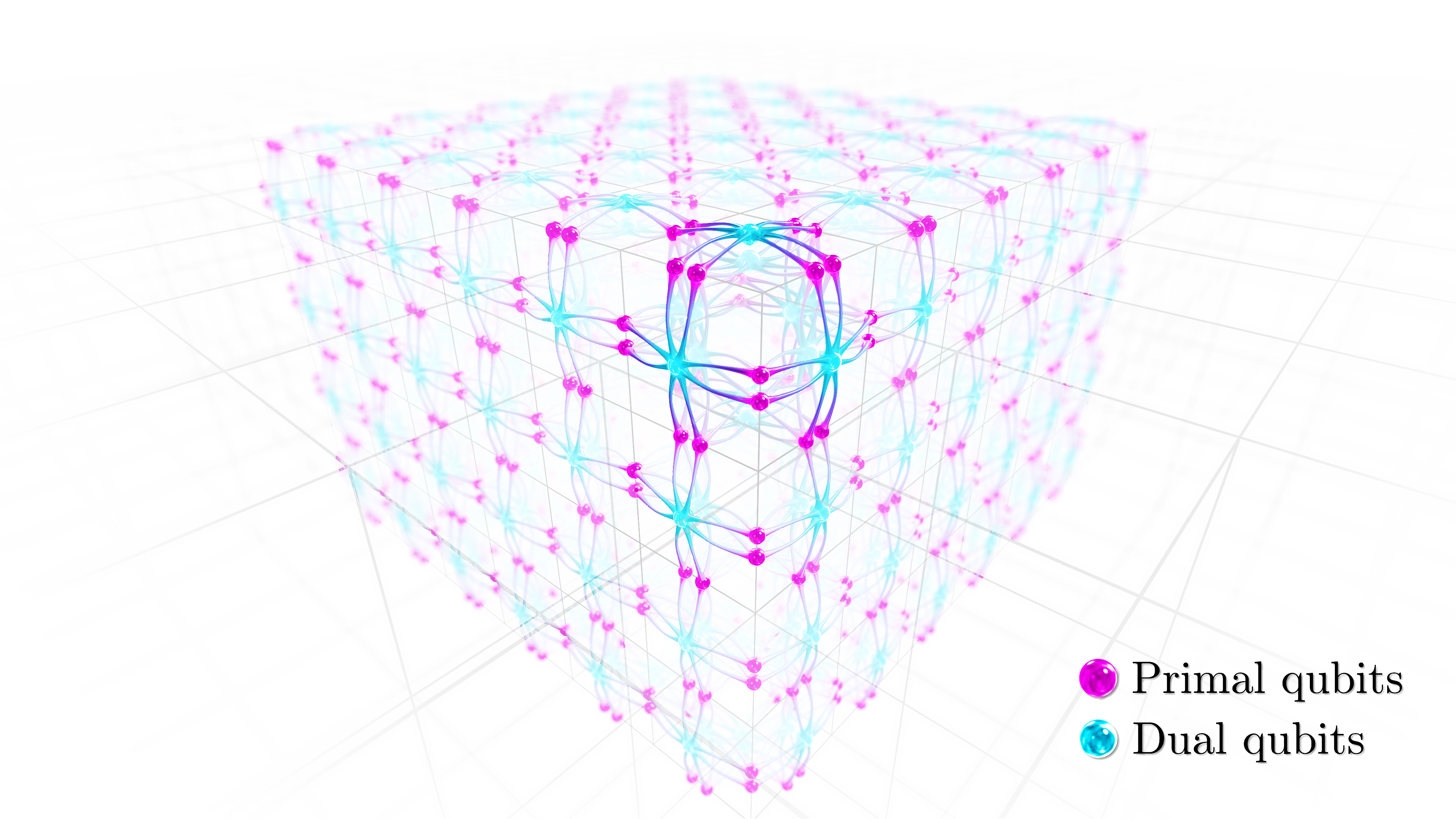}
\caption{Illustration of the cellular structure of the concatenated $\llbracket2,1,1\rrbracket$ lattice.}
\label{lattice}
\end{figure}

A state preparation for the logical $|\overline{+}\rangle$ state entails preparing the $|++\rangle$ state. Thus, no extra entangling gate is needed in the encoding procedure. The concatenated geometric structure, as shown in Figure~\ref{lattice}, is referred to as double-edge cubic lattice in Ref.~\cite{nickerson2018measurement} and is equivalent to the crazy-graph construction with $n=2$ of the bcc lattice~\cite{rudolph2017optimistic}. 

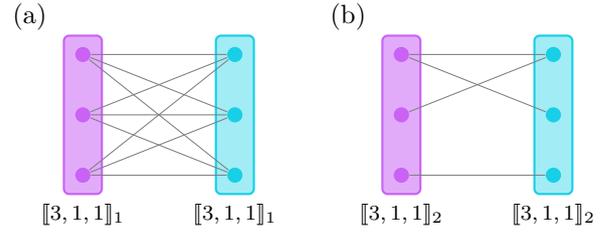
\begin{figure}[h]
\begin{subfigure}{0.47\columnwidth}
    \begin{tikzpicture}[baseline={([yshift=-.5ex]current bounding box.center)}]
    \draw[pink!75, fill=pink!40, rounded corners=2,thick]
     (-0.25, -0.25) rectangle (0.25,1.85) {};
    \draw[celeste, fill=celeste!40, rounded corners=2,thick]
     (1.75, -0.25) rectangle (2.25,1.85) {};
    \node[circle, fill=pink!75, inner sep=0pt, minimum size=0.2cm] (v1) at (0,0) {};
    \node[circle, fill=pink!75, inner sep=0pt, minimum size=0.2cm] (v2) at (0,0.8) {};
    \node[circle, fill=celeste, inner sep=0pt, minimum size=0.2cm] (u1) at (2,0) {};
    \node[circle, fill=celeste, inner sep=0pt, minimum size=0.2cm] (u2) at (2,0.8) {};
    \node[circle, fill=pink!75, inner sep=0pt, minimum size=0.2cm] (v3) at (0,1.6) {};
    \node[circle, fill=celeste, inner sep=0pt, minimum size=0.2cm] (u3) at (2,1.6) {};
    \node[below] at (0, -0.25) {\footnotesize $\llbracket3,1,1\rrbracket_1$};
    \node[below] at (2, -0.25) {\footnotesize $\llbracket3,1,1\rrbracket_1$};
    \node[above] at (-0.7, 1.8) {(a)};

    \draw[gray] (v1) -- (u1);
    \draw[gray] (v2) -- (u2);
    \draw[gray] (v1) -- (u3);
    \draw[gray] (v1) -- (u2);
    \draw[gray] (v2) -- (u1);
    \draw[gray] (v2) -- (u3);
    \draw[gray] (v3) -- (u2);
    \draw[gray] (v3) -- (u3);
    \draw[gray] (v3) -- (u1);
    \end{tikzpicture}
\phantomcaption\label{fig:logicalCZa}
\end{subfigure}\quad
\begin{subfigure}{0.47\columnwidth}
    \begin{tikzpicture}[baseline={([yshift=-.5ex]current bounding box.center)}]
    \draw[pink!75, fill=pink!40, rounded corners=2,thick]
     (-0.25, -0.25) rectangle (0.25,1.85) {};
    \draw[celeste, fill=celeste!40, rounded corners=2,thick]
     (1.75, -0.25) rectangle (2.25,1.85) {};
    \node[circle, fill=pink!75, inner sep=0pt, minimum size=0.2cm] (v1) at (0,0) {};
    \node[circle, fill=pink!75, inner sep=0pt, minimum size=0.2cm] (v2) at (0,0.8) {};
    \node[circle, fill=celeste, inner sep=0pt, minimum size=0.2cm] (u1) at (2,0) {};
    \node[circle, fill=celeste, inner sep=0pt, minimum size=0.2cm] (u2) at (2,0.8) {};
    \node[circle, fill=pink!75, inner sep=0pt, minimum size=0.2cm] (v3) at (0,1.6) {};
    \node[circle, fill=celeste, inner sep=0pt, minimum size=0.2cm] (u3) at (2,1.6) {};
    \node[below] at (0, -0.25) {\footnotesize $\llbracket3,1,1\rrbracket_2$};
    \node[below] at (2, -0.25) {\footnotesize $\llbracket3,1,1\rrbracket_2$};
    \node[above] at (-0.7, 1.8) {(b)};

    \draw[gray] (v1) -- (u1);
    \draw[gray] (v2) -- (u3);
    \draw[gray] (v3) -- (u2);
    \draw[gray] (v3) -- (u3);
    \end{tikzpicture}
    \phantomcaption\label{fig:logicalCZb}
\end{subfigure}
\caption{\subref{fig:logicalCZa} Logical $\overline{CZ}$ gates for the $\llbracket3,1,1\rrbracket_1$ code, \subref{fig:logicalCZb} logical $\overline{CZ}$ gates for the $\llbracket3,1,1\rrbracket_2$ code}.
\label{fig:logicalCZ}
\end{figure}

For $\llbracket3,1,1\rrbracket$ codes, two different constructions exist. The $\llbracket3,1,1\rrbracket_1$ code is the three-qubit repetition code.  The $\overline{CZ}$ gate can be performed with nine applications of physical $CZ$ gates~\cite{stephens2013high}, as shown in Figure~\ref{fig:logicalCZa}. 

For larger repetition codes, we could in principle perform a $\overline{CZ}$ gate by applying $O(n^2)$ $CZ$ gates. However, we expect the threshold to decrease as $n$ increases. Our numerical simulation shows that, indeed, $\llbracket3,1,1\rrbracket_1$ code performs worse than the $\llbracket2,1,1\rrbracket$ code.

The $\llbracket3,1,1\rrbracket_2$ code can be viewed as an ``interpolation'' between the $\llbracket2,1,1\rrbracket$ code and the $\llbracket4,1,1,2\rrbracket$ code, which is manifested by its logical operators defined by $\overline{Z} = Z_1Z_3$ and $\overline{X} = X_1$. The logical $\overline{CZ}$ gate of this code is shown in Figure~\ref{fig:logicalCZb}.

Because none of the codes mentioned in this section can detect an arbitrary single-qubit error, it is a priori not obvious why there should be a $\mathcal{C}$-detectable state preparation procedure. We claim below that this is nevertheless the case, by showing explicit examples. Consider the gate sequence in Figure~\ref{fig:sequence}. In the ensuing discussion we shall refer to the individual code blocks as central (C), northern (N), southern (S), western (W), and eastern (E), depending on their relative location with respect to the central block. Since every code block can be interpreted as a central block and the scheduling in Figure~\ref{fig:sequence} can be applied in a uniform way throughout the entire lattice, it suffices to consider single-qubit errors occurring on the central block and two-qubit errors that can occur after each $CZ$ gate.

Let us first consider single-qubit errors. Since $CZ$ gates commute with $Z$ errors, $Z$ errors do not propagate and remain $\mathcal{C}$-detectable. For a single-qubit $X$ error occurring in the central block, we make the following observation: if such an error occurs after Step 4, it propagates to a tensor product of single-qubit $Z$ errors and identity on the N, S, E, and W code blocks. Therefore, in this case, the error propagates to a $\mathcal{C}$-detectable error. On the other hand, if an error occurs between Step 1 and Step 4, it propagates to a tensor product of single or two-qubit $Z$ errors within each block. Although a two-qubit error within a block is a logical $Z$ error of that code, the propagated error is still $\mathcal{C}$-detectable because the resulting error is equivalent, up to an outer code stabilizer, to a tensor product of single-qubit $Z$ errors and identity on each code block. Specifically, the stabilizer is given by $\overline{X}_C \overline{Z}_N \overline{Z}_S \overline{Z}_W \overline{Z}_E$.

Moreover, all the two-qubit circuit-level errors propagate to $\llbracket2,1,1\rrbracket$-detectable errors. We have investigated how these errors propagate, up to stabilizers, the result of which is summarized in Appendix~\ref{appendix:a}. We also note that the schemes based on the $\llbracket3,1,1\rrbracket_1$ and the $\llbracket3,1,1\rrbracket_2$ code are $\mathcal{C}$-detectable; see Appendix~\ref{appendix:a}.

\begin{figure}[h]
\centering
	\begin{tikzpicture}[scale=0.6]
		\draw[pink!75, fill=pink!40, rounded corners=2,thick]
		 (-1.25, -0.25) rectangle (-0.75,0.75) {};
		 \draw[pink!75, fill=pink!40, rounded corners=2,thick]
		 (-0.5, 1) rectangle (0.5,1.5) {};
		 \draw[pink!75, fill=pink!40, rounded corners=2,thick]
		 (0.75, -0.25) rectangle (1.25,0.75) {};
		 \draw[pink!75, fill=pink!40, rounded corners=2,thick]
		 (-0.5, -1) rectangle (0.5,-0.5) {};
		\draw[celeste, fill=celeste!40, rounded corners=2,thick]
		 (-0.25, -0.25) rectangle (0.25,0.75) {};
		\node[circle, fill=celeste, inner sep=0pt, minimum size=0.12cm] (v1c) at (0,0) {};
		\node[circle, fill=celeste, inner sep=0pt, minimum size=0.12cm] (v2c) at (0,0.5) {};
		
		\node[circle, fill=pink!75, inner sep=0pt, minimum size=0.12cm] (v1e) at (1,0) {};
		\node[circle, fill=pink!75, inner sep=0pt, minimum size=0.12cm] (v2e) at (1,0.5) {};
		
		\node[circle, fill=pink!75, inner sep=0pt, minimum size=0.12cm] (v1w) at (-1,0) {};
		\node[circle, fill=pink!75, inner sep=0pt, minimum size=0.12cm] (v2w) at (-1,0.5) {};
		
		\node[circle, fill=pink!75, inner sep=0pt, minimum size=0.12cm] (v1n) at (-0.25,1.25) {};
		\node[circle, fill=pink!75, inner sep=0pt, minimum size=0.12cm] (v2n) at (0.25,1.25) {};
		
		\node[circle, fill=pink!75, inner sep=0pt, minimum size=0.12cm] (v1s) at (-0.25,-0.75) {};
		\node[circle, fill=pink!75, inner sep=0pt, minimum size=0.12cm] (v2s) at (0.25,-0.75) {};
		\draw[gray] (v1c) -- (v1w);
		\draw[gray] (v2c) -- (v2w);
		\node[] () at (0, -1.5) {\footnotesize Step $1$};
	\end{tikzpicture}
	$\quad$
	\begin{tikzpicture}[scale=0.6]
		\draw[pink!75, fill=pink!40, rounded corners=2,thick]
		 (-1.25, -0.25) rectangle (-0.75,0.75) {};
		 \draw[pink!75, fill=pink!40, rounded corners=2,thick]
		 (-0.5, 1) rectangle (0.5,1.5) {};
		 \draw[pink!75, fill=pink!40, rounded corners=2,thick]
		 (0.75, -0.25) rectangle (1.25,0.75) {};
		 \draw[pink!75, fill=pink!40, rounded corners=2,thick]
		 (-0.5, -1) rectangle (0.5,-0.5) {};
		\draw[celeste, fill=celeste!40, rounded corners=2,thick]
		 (-0.25, -0.25) rectangle (0.25,0.75) {};
		\node[circle, fill=celeste, inner sep=0pt, minimum size=0.12cm] (v1c) at (0,0) {};
		\node[circle, fill=celeste, inner sep=0pt, minimum size=0.12cm] (v2c) at (0,0.5) {};
		
		\node[circle, fill=pink!75, inner sep=0pt, minimum size=0.12cm] (v1e) at (1,0) {};
		\node[circle, fill=pink!75, inner sep=0pt, minimum size=0.12cm] (v2e) at (1,0.5) {};
		
		\node[circle, fill=pink!75, inner sep=0pt, minimum size=0.12cm] (v1w) at (-1,0) {};
		\node[circle, fill=pink!75, inner sep=0pt, minimum size=0.12cm] (v2w) at (-1,0.5) {};
		
		\node[circle, fill=pink!75, inner sep=0pt, minimum size=0.12cm] (v1n) at (-0.25,1.25) {};
		\node[circle, fill=pink!75, inner sep=0pt, minimum size=0.12cm] (v2n) at (0.25,1.25) {};
		
		\node[circle, fill=pink!75, inner sep=0pt, minimum size=0.12cm] (v1s) at (-0.25,-0.75) {};
		\node[circle, fill=pink!75, inner sep=0pt, minimum size=0.12cm] (v2s) at (0.25,-0.75) {};
		\draw[gray] (v1c) -- (v2w);
		\draw[gray] (v2c) -- (v1w);
		\node[] () at (0, -1.5) {\footnotesize Step $2$};
	\end{tikzpicture}
	$\quad$
	\begin{tikzpicture}[scale=0.6]
		\draw[pink!75, fill=pink!40, rounded corners=2,thick]
		 (-1.25, -0.25) rectangle (-0.75,0.75) {};
		 \draw[pink!75, fill=pink!40, rounded corners=2,thick]
		 (-0.5, 1) rectangle (0.5,1.5) {};
		 \draw[pink!75, fill=pink!40, rounded corners=2,thick]
		 (0.75, -0.25) rectangle (1.25,0.75) {};
		 \draw[pink!75, fill=pink!40, rounded corners=2,thick]
		 (-0.5, -1) rectangle (0.5,-0.5) {};
		\draw[celeste, fill=celeste!40, rounded corners=2,thick]
		 (-0.25, -0.25) rectangle (0.25,0.75) {};
		\node[circle, fill=celeste, inner sep=0pt, minimum size=0.12cm] (v1c) at (0,0) {};
		\node[circle, fill=celeste, inner sep=0pt, minimum size=0.12cm] (v2c) at (0,0.5) {};
		
		\node[circle, fill=pink!75, inner sep=0pt, minimum size=0.12cm] (v1e) at (1,0) {};
		\node[circle, fill=pink!75, inner sep=0pt, minimum size=0.12cm] (v2e) at (1,0.5) {};
		
		\node[circle, fill=pink!75, inner sep=0pt, minimum size=0.12cm] (v1w) at (-1,0) {};
		\node[circle, fill=pink!75, inner sep=0pt, minimum size=0.12cm] (v2w) at (-1,0.5) {};
		
		\node[circle, fill=pink!75, inner sep=0pt, minimum size=0.12cm] (v1n) at (-0.25,1.25) {};
		\node[circle, fill=pink!75, inner sep=0pt, minimum size=0.12cm] (v2n) at (0.25,1.25) {};
		
		\node[circle, fill=pink!75, inner sep=0pt, minimum size=0.12cm] (v1s) at (-0.25,-0.75) {};
		\node[circle, fill=pink!75, inner sep=0pt, minimum size=0.12cm] (v2s) at (0.25,-0.75) {};
		\draw[gray] (v1c) -- (v1e);
		\draw[gray] (v2c) -- (v2e);
		\node[] () at (0, -1.5) {\footnotesize Step $3$};
	\end{tikzpicture}
	$\quad$
	\begin{tikzpicture}[scale=0.6]
		\draw[pink!75, fill=pink!40, rounded corners=2,thick]
		 (-1.25, -0.25) rectangle (-0.75,0.75) {};
		 \draw[pink!75, fill=pink!40, rounded corners=2,thick]
		 (-0.5, 1) rectangle (0.5,1.5) {};
		 \draw[pink!75, fill=pink!40, rounded corners=2,thick]
		 (0.75, -0.25) rectangle (1.25,0.75) {};
		 \draw[pink!75, fill=pink!40, rounded corners=2,thick]
		 (-0.5, -1) rectangle (0.5,-0.5) {};
		\draw[celeste, fill=celeste!40, rounded corners=2,thick]
		 (-0.25, -0.25) rectangle (0.25,0.75) {};
		\node[circle, fill=celeste, inner sep=0pt, minimum size=0.12cm] (v1c) at (0,0) {};
		\node[circle, draw=celeste, fill=celeste!40, inner sep=0pt, minimum size=0.18cm] (v2c) at (0,0.5) {};
		\node[text=gray] () at (0,0.5) {\tiny $X$};
		
		\node[circle, draw=pink!75, fill=pink!40, inner sep=0pt, minimum size=0.18cm] (v1e) at (1,0) {};
		\node[text=gray] () at (1,0) {\tiny $Z$};
		\node[circle, fill=pink!75, inner sep=0pt, minimum size=0.12cm] (v2e) at (1,0.5) {};
		
		\node[circle, fill=pink!75, inner sep=0pt, minimum size=0.12cm] (v1w) at (-1,0) {};
		\node[circle, fill=pink!75, inner sep=0pt, minimum size=0.12cm] (v2w) at (-1,0.5) {};
		
		\node[circle, fill=pink!75, inner sep=0pt, minimum size=0.12cm] (v1n) at (-0.25,1.25) {};
		\node[circle, fill=pink!75, inner sep=0pt, minimum size=0.12cm] (v2n) at (0.25,1.25) {};
		
		\node[circle, fill=pink!75, inner sep=0pt, minimum size=0.12cm] (v1s) at (-0.25,-0.75) {};
		\node[circle, fill=pink!75, inner sep=0pt, minimum size=0.12cm] (v2s) at (0.25,-0.75) {};
		\draw[gray] (v1c) -- (v2e);
		\draw[gray] (v2c) -- (v1e);
		\node[] () at (0, -1.5) {\footnotesize Step $4$};
	\end{tikzpicture}\\
	\vspace{0.5cm}%
	\begin{tikzpicture}[scale=0.6]
		\draw[pink!75, fill=pink!40, rounded corners=2,thick]
		 (-1.25, -0.25) rectangle (-0.75,0.75) {};
		 \draw[pink!75, fill=pink!40, rounded corners=2,thick]
		 (-0.5, 1) rectangle (0.5,1.5) {};
		 \draw[pink!75, fill=pink!40, rounded corners=2,thick]
		 (0.75, -0.25) rectangle (1.25,0.75) {};
		 \draw[pink!75, fill=pink!40, rounded corners=2,thick]
		 (-0.5, -1) rectangle (0.5,-0.5) {};
		\draw[celeste, fill=celeste!40, rounded corners=2,thick]
		 (-0.25, -0.25) rectangle (0.25,0.75) {};
		\node[circle, fill=celeste, inner sep=0pt, minimum size=0.12cm] (v1c) at (0,0) {};
		\node[circle, draw=celeste, fill=celeste!40, inner sep=0pt, minimum size=0.18cm] (v2c) at (0,0.5) {};
		\node[text=gray] () at (0,0.5) {\tiny $X$};
		
		\node[circle, draw=pink!75, fill=pink!40, inner sep=0pt, minimum size=0.18cm] (v1e) at (1,0) {};
		\node[text=gray] () at (1,0) {\tiny $Z$};
		\node[circle, fill=pink!75, inner sep=0pt, minimum size=0.12cm] (v2e) at (1,0.5) {};
		
		\node[circle, fill=pink!75, inner sep=0pt, minimum size=0.12cm] (v1w) at (-1,0) {};
		\node[circle, fill=pink!75, inner sep=0pt, minimum size=0.12cm] (v2w) at (-1,0.5) {};
		
		\node[circle, fill=pink!75, inner sep=0pt, minimum size=0.12cm] (v1n) at (-0.25,1.25) {};
		\node[circle, fill=pink!75, inner sep=0pt, minimum size=0.12cm] (v2n) at (0.25,1.25) {};
		
		\node[circle, fill=pink!75, inner sep=0pt, minimum size=0.12cm] (v1s) at (-0.25,-0.75) {};
		\node[circle, draw=pink!75, fill=pink!40, inner sep=0pt, minimum size=0.18cm] (v2s) at (0.25,-0.75) {};
		\node[text=gray] () at (0.25,-0.75) {\tiny $Z$};
		\draw[gray] (v1c) -- (v1s);
		\draw[gray] (v2c) -- (v2s);
		\node[] () at (0, -1.5) {\footnotesize Step $5$};
	\end{tikzpicture}
	$\quad$
	\begin{tikzpicture}[scale=0.6]
		\draw[pink!75, fill=pink!40, rounded corners=2,thick]
		 (-1.25, -0.25) rectangle (-0.75,0.75) {};
		 \draw[pink!75, fill=pink!40, rounded corners=2,thick]
		 (-0.5, 1) rectangle (0.5,1.5) {};
		 \draw[pink!75, fill=pink!40, rounded corners=2,thick]
		 (0.75, -0.25) rectangle (1.25,0.75) {};
		 \draw[pink!75, fill=pink!40, rounded corners=2,thick]
		 (-0.5, -1) rectangle (0.5,-0.5) {};
		\draw[celeste, fill=celeste!40, rounded corners=2,thick]
		 (-0.25, -0.25) rectangle (0.25,0.75) {};
		\node[circle, fill=celeste, inner sep=0pt, minimum size=0.12cm] (v1c) at (0,0) {};
		\node[circle, draw=celeste, fill=celeste!40, inner sep=0pt, minimum size=0.18cm] (v2c) at (0,0.5) {};
		\node[text=gray] () at (0,0.5) {\tiny $X$};
		
		\node[circle, draw=pink!75, fill=pink!40, inner sep=0pt, minimum size=0.18cm] (v1e) at (1,0) {};
		\node[text=gray] () at (1,0) {\tiny $Z$};
		\node[circle, fill=pink!75, inner sep=0pt, minimum size=0.12cm] (v2e) at (1,0.5) {};
		
		\node[circle, fill=pink!75, inner sep=0pt, minimum size=0.12cm] (v1w) at (-1,0) {};
		\node[circle, fill=pink!75, inner sep=0pt, minimum size=0.12cm] (v2w) at (-1,0.5) {};
		
		\node[circle, fill=pink!75, inner sep=0pt, minimum size=0.12cm] (v1n) at (-0.25,1.25) {};
		\node[circle, fill=pink!75, inner sep=0pt, minimum size=0.12cm] (v2n) at (0.25,1.25) {};
		
		\node[circle, draw=pink!75, fill=pink!40, inner sep=0pt, minimum size=0.18cm] (v1s) at (-0.25,-0.75) {};
		 \node[text=gray] () at (-0.25,-0.75) {\tiny $Z$};
		\node[circle, draw=pink!75, fill=pink!40, inner sep=0pt, minimum size=0.18cm] (v2s) at (0.25,-0.75) {};
		\node[text=gray] () at (0.25,-0.75) {\tiny $Z$};
		\draw[gray] (v1c) -- (v2s);
		\draw[gray] (v2c) -- (v1s);
		\node[] () at (0, -1.5) {\footnotesize Step $6$};
	\end{tikzpicture}
	$\quad$
	\begin{tikzpicture}[scale=0.6]
		\draw[pink!75, fill=pink!40, rounded corners=2,thick]
		 (-1.25, -0.25) rectangle (-0.75,0.75) {};
		 \draw[pink!75, fill=pink!40, rounded corners=2,thick]
		 (-0.5, 1) rectangle (0.5,1.5) {};
		 \draw[pink!75, fill=pink!40, rounded corners=2,thick]
		 (0.75, -0.25) rectangle (1.25,0.75) {};
		 \draw[pink!75, fill=pink!40, rounded corners=2,thick]
		 (-0.5, -1) rectangle (0.5,-0.5) {};
		\draw[celeste, fill=celeste!40, rounded corners=2,thick]
		 (-0.25, -0.25) rectangle (0.25,0.75) {};
		\node[circle, fill=celeste, inner sep=0pt, minimum size=0.12cm] (v1c) at (0,0) {};
		\draw[gray] (v1c) -- (v1n);
		\node[circle, fill=pink!75, inner sep=0pt, minimum size=0.12cm] (v1n) at (-0.25,1.25) {};
		\node[circle, draw=pink!75, fill=pink!40, inner sep=0pt, minimum size=0.18cm] (v2n) at (0.25,1.25) {};
		\node[text=gray] () at (0.25,1.25) {\tiny $Z$};
		\node[circle, draw=celeste, fill=celeste!40, inner sep=0pt, minimum size=0.18cm] (v2c) at (0,0.5) {};
		\node[text=gray] () at (0,0.5) {\tiny $X$};
		\draw[gray] (v2c) -- (v2n);
		
		\node[circle, draw=pink!75, fill=pink!40, inner sep=0pt, minimum size=0.18cm] (v1e) at (1,0) {};
		\node[text=gray] () at (1,0) {\tiny $Z$};
		\node[circle, fill=pink!75, inner sep=0pt, minimum size=0.12cm] (v2e) at (1,0.5) {};
		
		\node[circle, fill=pink!75, inner sep=0pt, minimum size=0.12cm] (v1w) at (-1,0) {};
		\node[circle, fill=pink!75, inner sep=0pt, minimum size=0.12cm] (v2w) at (-1,0.5) {};
		
		\node[circle, draw=pink!75, fill=pink!40, inner sep=0pt, minimum size=0.18cm] (v1s) at (-0.25,-0.75) {};
		\node[text=gray] () at (-0.25,-0.75) {\tiny $Z$};
		\node[circle, draw=pink!75, fill=pink!40, inner sep=0pt, minimum size=0.18cm] (v2s) at (0.25,-0.75) {};
		\node[text=gray] () at (0.25,-0.75) {\tiny $Z$};
		\node[] () at (0, -1.5) {\footnotesize Step $7$};
	\end{tikzpicture}
	$\quad$
	\begin{tikzpicture}[scale=0.6]
		\draw[pink!75, fill=pink!40, rounded corners=2,thick]
		 (-1.25, -0.25) rectangle (-0.75,0.75) {};
		 \draw[pink!75, fill=pink!40, rounded corners=2,thick]
		 (-0.5, 1) rectangle (0.5,1.5) {};
		 \draw[pink!75, fill=pink!40, rounded corners=2,thick]
		 (0.75, -0.25) rectangle (1.25,0.75) {};
		 \draw[pink!75, fill=pink!40, rounded corners=2,thick]
		 (-0.5, -1) rectangle (0.5,-0.5) {};
		\draw[celeste, fill=celeste!40, rounded corners=2,thick]
		 (-0.25, -0.25) rectangle (0.25,0.75) {};
		\node[circle, fill=celeste, inner sep=0pt, minimum size=0.12cm] (v1c) at (0,0) {};
		\node[circle, draw=pink!75, fill=pink!40, inner sep=0pt, minimum size=0.18cm] (v2n) at (0.25,1.25) {};
		\node[text=gray] () at (0.25,1.25) {\tiny $Z$};
		\draw[gray] (v1c) -- (v2n);
		\draw[gray] (v2c) -- (v1n);
		
		\node[circle, draw=celeste, fill=celeste!40, inner sep=0pt, minimum size=0.18cm] (v2c) at (0,0.5) {};
		\node[text=gray] () at (0,0.5) {\tiny $X$};
		\node[circle, draw=pink!75, fill=pink!40, inner sep=0pt, minimum size=0.18cm] (v1n) at (-0.25,1.25) {};
		\node[text=gray] () at (-0.25,1.25) {\tiny $Z$};
		
		\node[circle, draw=pink!75, fill=pink!40, inner sep=0pt, minimum size=0.18cm] (v1e) at (1,0) {};
		\node[text=gray] () at (1,0) {\tiny $Z$};
		\node[circle, fill=pink!75, inner sep=0pt, minimum size=0.12cm] (v2e) at (1,0.5) {};
		
		\node[circle, fill=pink!75, inner sep=0pt, minimum size=0.12cm] (v1w) at (-1,0) {};
		\node[circle, fill=pink!75, inner sep=0pt, minimum size=0.12cm] (v2w) at (-1,0.5) {};

		\node[circle, draw=pink!75, fill=pink!40, inner sep=0pt, minimum size=0.18cm] (v1s) at (-0.25,-0.75) {};
		\node[text=gray] () at (-0.25,-0.75) {\tiny $Z$};
		\node[circle, draw=pink!75, fill=pink!40, inner sep=0pt, minimum size=0.18cm] (v2s) at (0.25,-0.75) {};
		\node[text=gray] () at (0.25,-0.75) {\tiny $Z$};
		\node[] () at (0, -1.5) {\footnotesize Step $8$};
	\end{tikzpicture}
		\caption{``Natural" gate sequence of the $\llbracket2,1,1\rrbracket$ scheme that fails to be $\mathcal{C}$-detectable. Note that a single-qubit $X$ error happening on one of the central qubits could propagate to two undetectable errors in the N and S blocks.
	\label{fig:sequence_bad}}
	\end{figure}
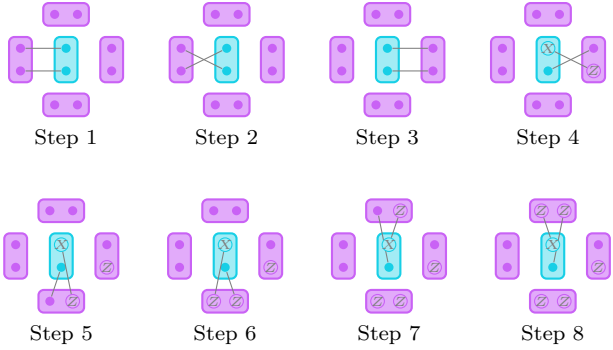

We emphasize that the ordering of the gates in Figure~\ref{fig:sequence} is important. This gate sequence is in particular different from the ``natural" gate sequence proposed in Ref.~\cite{stephens2013high} that applies all physical $CZ$ gates of a logical $\overline{CZ}$ gate consecutively (shown in Figure~\ref{fig:sequence_bad}). As one can see in Figure~\ref{fig:sequence_bad}, there are circuit-level errors which propagate to errors that are \emph{not} $\llbracket2,1,1\rrbracket$-detectable. Thus, the fact that all circuit-level errors are $\llbracket2,1,1\rrbracket$-detectable is a special property of the circuit shown in Figure~\ref{fig:sequence}. The inner codes and their respective stabilizers and logical operators are summarized in Table~\ref{tab:stabs}.

\begin{figure}[h]
\centering
\begin{tikzpicture}[scale=0.6]
    \draw[pink!75, fill=pink!40, rounded corners=2,thick]
     (-1.25, -0.25) rectangle (-0.75,0.75) {};
      \node[below] () at (-1.55, 0.53) {\tiny $W$};
     \draw[pink!75, fill=pink!40, rounded corners=2,thick]
     (-0.5, 1) rectangle (0.5,1.5) {};
       \node[below] () at (-0.75, 1.56) {\tiny $N$};
     \draw[pink!75, fill=pink!40, rounded corners=2,thick]
     (0.75, -0.25) rectangle (1.25,0.75) {};
      \node[below] () at (0.55, 0.53) {\tiny $E$};
     \draw[pink!75, fill=pink!40, rounded corners=2,thick]
     (-0.5, -1) rectangle (0.5,-0.4) {};
      \node[below] () at (-0.75, -0.44) {\tiny $S$};
    \draw[celeste, fill=celeste!40, rounded corners=2,thick]
     (-0.25, -0.25) rectangle (0.25,0.75) {};
      \node[below] () at (-0.5, 0.53) {\tiny $C$};
    \node[circle, fill=celeste, inner sep=0pt, minimum size=0.12cm] (v1c) at (0,0) {};
    \node[circle, fill=celeste, inner sep=0pt, minimum size=0.12cm] (v2c) at (0,0.5) {};
    
    \node[circle, fill=pink!75, inner sep=0pt, minimum size=0.12cm] (v1e) at (1,0) {};
    \node[circle, fill=pink!75, inner sep=0pt, minimum size=0.12cm] (v2e) at (1,0.5) {};
    
    \node[circle, fill=pink!75, inner sep=0pt, minimum size=0.12cm] (v1w) at (-1,0) {};
    \node[circle, fill=pink!75, inner sep=0pt, minimum size=0.12cm] (v2w) at (-1,0.5) {};
    
    \node[circle, fill=pink!75, inner sep=0pt, minimum size=0.12cm] (v1n) at (-0.25,1.25) {};
    \node[circle, fill=pink!75, inner sep=0pt, minimum size=0.12cm] (v2n) at (0.25,1.25) {};
    
    \node[circle, fill=pink!75, inner sep=0pt, minimum size=0.12cm] (v1s) at (-0.25,-0.75) {};
    \node[circle, fill=pink!75, inner sep=0pt, minimum size=0.12cm] (v2s) at (0.25,-0.75) {};
    \draw[gray] (v1c) -- (v1w);
    \draw[gray] (v2c) -- (v2w);
    \node[] () at (0, -1.5) {\footnotesize Step $1$};
\end{tikzpicture}\;\;\;
\begin{tikzpicture}[scale=0.6]
    \draw[pink!75, fill=pink!40, rounded corners=2,thick]
     (-1.25, -0.25) rectangle (-0.75,0.75) {};
     \draw[pink!75, fill=pink!40, rounded corners=2,thick]
     (-0.5, 1) rectangle (0.5,1.5) {};
     \draw[pink!75, fill=pink!40, rounded corners=2,thick]
     (0.75, -0.25) rectangle (1.25,0.75) {};
     \draw[pink!75, fill=pink!40, rounded corners=2,thick]
     (-0.5, -1) rectangle (0.5,-0.5) {};
    \draw[celeste, fill=celeste!40, rounded corners=2,thick]
     (-0.25, -0.25) rectangle (0.25,0.75) {};
    \node[circle, fill=celeste, inner sep=0pt, minimum size=0.12cm] (v1c) at (0,0) {};
    \node[circle, fill=celeste, inner sep=0pt, minimum size=0.12cm] (v2c) at (0,0.5) {};
    
    \node[circle, fill=pink!75, inner sep=0pt, minimum size=0.12cm] (v1e) at (1,0) {};
    \node[circle, fill=pink!75, inner sep=0pt, minimum size=0.12cm] (v2e) at (1,0.5) {};
    
    \node[circle, fill=pink!75, inner sep=0pt, minimum size=0.12cm] (v1w) at (-1,0) {};
    \node[circle, fill=pink!75, inner sep=0pt, minimum size=0.12cm] (v2w) at (-1,0.5) {};
    
    \node[circle, fill=pink!75, inner sep=0pt, minimum size=0.12cm] (v1n) at (-0.25,1.25) {};
    \node[circle, fill=pink!75, inner sep=0pt, minimum size=0.12cm] (v2n) at (0.25,1.25) {};
    
    \node[circle, fill=pink!75, inner sep=0pt, minimum size=0.12cm] (v1s) at (-0.25,-0.75) {};
    \node[circle, fill=pink!75, inner sep=0pt, minimum size=0.12cm] (v2s) at (0.25,-0.75) {};
    \draw[gray] (v1c) -- (v1e);
    \draw[gray] (v2c) -- (v2e);
    \node[] () at (0, -1.5) {\footnotesize Step $2$};
\end{tikzpicture}
$\quad$
\begin{tikzpicture}[scale=0.6]
    \draw[pink!75, fill=pink!40, rounded corners=2,thick]
     (-1.25, -0.25) rectangle (-0.75,0.75) {};
     \draw[pink!75, fill=pink!40, rounded corners=2,thick]
     (-0.5, 1) rectangle (0.5,1.5) {};
     \draw[pink!75, fill=pink!40, rounded corners=2,thick]
     (0.75, -0.25) rectangle (1.25,0.75) {};
     \draw[pink!75, fill=pink!40, rounded corners=2,thick]
     (-0.5, -1) rectangle (0.5,-0.5) {};
    \draw[celeste, fill=celeste!40, rounded corners=2,thick]
     (-0.25, -0.25) rectangle (0.25,0.75) {};
    \node[circle, fill=celeste, inner sep=0pt, minimum size=0.12cm] (v1c) at (0,0) {};
    \node[circle, fill=celeste, inner sep=0pt, minimum size=0.12cm] (v2c) at (0,0.5) {};
    
    \node[circle, fill=pink!75, inner sep=0pt, minimum size=0.12cm] (v1e) at (1,0) {};
    \node[circle, fill=pink!75, inner sep=0pt, minimum size=0.12cm] (v2e) at (1,0.5) {};
    
    \node[circle, fill=pink!75, inner sep=0pt, minimum size=0.12cm] (v1w) at (-1,0) {};
    \node[circle, fill=pink!75, inner sep=0pt, minimum size=0.12cm] (v2w) at (-1,0.5) {};
    
    \node[circle, fill=pink!75, inner sep=0pt, minimum size=0.12cm] (v1n) at (-0.25,1.25) {};
    \node[circle, fill=pink!75, inner sep=0pt, minimum size=0.12cm] (v2n) at (0.25,1.25) {};
    
    \node[circle, fill=pink!75, inner sep=0pt, minimum size=0.12cm] (v1s) at (-0.25,-0.75) {};
    \node[circle, fill=pink!75, inner sep=0pt, minimum size=0.12cm] (v2s) at (0.25,-0.75) {};
    \draw[gray] (v1c) -- (v1s);
    \draw[gray] (v2c) -- (v2s);
    \node[] () at (0, -1.5) {\footnotesize Step $3$};
\end{tikzpicture}
$\quad$
\begin{tikzpicture}[scale=0.6]
    \draw[pink!75, fill=pink!40, rounded corners=2,thick]
     (-1.25, -0.25) rectangle (-0.75,0.75) {};
     \draw[pink!75, fill=pink!40, rounded corners=2,thick]
     (-0.5, 1) rectangle (0.5,1.5) {};
     \draw[pink!75, fill=pink!40, rounded corners=2,thick]
     (0.75, -0.25) rectangle (1.25,0.75) {};
     \draw[pink!75, fill=pink!40, rounded corners=2,thick]
     (-0.5, -1) rectangle (0.5,-0.5) {};
    \draw[celeste, fill=celeste!40, rounded corners=2,thick]
     (-0.25, -0.25) rectangle (0.25,0.75) {};
    \node[circle, fill=celeste, inner sep=0pt, minimum size=0.12cm] (v1c) at (0,0) {};
    \draw[gray] (v1c) -- (v1n);
    \node[circle, fill=pink!75, inner sep=0pt, minimum size=0.12cm] (v1n) at (-0.25,1.25) {};
    \node[circle, fill=pink!75, inner sep=0pt, minimum size=0.12cm] (v2n) at (0.25,1.25) {};
    \node[circle, fill=celeste, inner sep=0pt, minimum size=0.12cm] (v2c) at (0,0.5) {};
    \draw[gray] (v2c) -- (v2n);
    
    \node[circle, fill=pink!75, inner sep=0pt, minimum size=0.12cm] (v1e) at (1,0) {};
    \node[circle, fill=pink!75, inner sep=0pt, minimum size=0.12cm] (v2e) at (1,0.5) {};
    
    \node[circle, fill=pink!75, inner sep=0pt, minimum size=0.12cm] (v1w) at (-1,0) {};
    \node[circle, fill=pink!75, inner sep=0pt, minimum size=0.12cm] (v2w) at (-1,0.5) {};
    
    \node[circle, fill=pink!75, inner sep=0pt, minimum size=0.12cm] (v1s) at (-0.25,-0.75) {};
    \node[circle, fill=pink!75, inner sep=0pt, minimum size=0.12cm] (v2s) at (0.25,-0.75) {};
    \node[] () at (0, -1.5) {\footnotesize Step $4$};
\end{tikzpicture}\\
\vspace{0.5cm}
\begin{tikzpicture}[scale=0.6]
    \draw[pink!75, fill=pink!40, rounded corners=2,thick]
     (-1.25, -0.25) rectangle (-0.75,0.75) {};
     \draw[pink!75, fill=pink!40, rounded corners=2,thick]
     (-0.5, 1) rectangle (0.5,1.5) {};
     \draw[pink!75, fill=pink!40, rounded corners=2,thick]
     (0.75, -0.25) rectangle (1.25,0.75) {};
     \draw[pink!75, fill=pink!40, rounded corners=2,thick]
     (-0.5, -1) rectangle (0.5,-0.5) {};
    \draw[celeste, fill=celeste!40, rounded corners=2,thick]
     (-0.25, -0.25) rectangle (0.25,0.75) {};
    \node[circle, fill=celeste, inner sep=0pt, minimum size=0.12cm] (v1c) at (0,0) {};
    \node[circle, fill=celeste, inner sep=0pt, minimum size=0.12cm] (v2c) at (0,0.5) {};
    
    \node[circle, fill=pink!75, inner sep=0pt, minimum size=0.12cm] (v1e) at (1,0) {};
    \node[circle, fill=pink!75, inner sep=0pt, minimum size=0.12cm] (v2e) at (1,0.5) {};
    
    \node[circle, fill=pink!75, inner sep=0pt, minimum size=0.12cm] (v1w) at (-1,0) {};
    \node[circle, fill=pink!75, inner sep=0pt, minimum size=0.12cm] (v2w) at (-1,0.5) {};
    
    \node[circle, fill=pink!75, inner sep=0pt, minimum size=0.12cm] (v1n) at (-0.25,1.25) {};
    \node[circle, fill=pink!75, inner sep=0pt, minimum size=0.12cm] (v2n) at (0.25,1.25) {};
    
    \node[circle, fill=pink!75, inner sep=0pt, minimum size=0.12cm] (v1s) at (-0.25,-0.75) {};
    \node[circle, fill=pink!75, inner sep=0pt, minimum size=0.12cm] (v2s) at (0.25,-0.75) {};
    \draw[gray] (v1c) -- (v2w);
    \draw[gray] (v2c) -- (v1w);
    \node[] () at (0, -1.5) {\footnotesize Step $5$};
\end{tikzpicture}
$\quad$
\begin{tikzpicture}[scale=0.6]
    \draw[pink!75, fill=pink!40, rounded corners=2,thick]
     (-1.25, -0.25) rectangle (-0.75,0.75) {};
     \draw[pink!75, fill=pink!40, rounded corners=2,thick]
     (-0.5, 1) rectangle (0.5,1.5) {};
     \draw[pink!75, fill=pink!40, rounded corners=2,thick]
     (0.75, -0.25) rectangle (1.25,0.75) {};
     \draw[pink!75, fill=pink!40, rounded corners=2,thick]
     (-0.5, -1) rectangle (0.5,-0.5) {};
    \draw[celeste, fill=celeste!40, rounded corners=2,thick]
     (-0.25, -0.25) rectangle (0.25,0.75) {};
    \node[circle, fill=celeste, inner sep=0pt, minimum size=0.12cm] (v1c) at (0,0) {};
    \node[circle, fill=celeste, inner sep=0pt, minimum size=0.12cm] (v2c) at (0,0.5) {};
    
    \node[circle, fill=pink!75, inner sep=0pt, minimum size=0.12cm] (v1e) at (1,0) {};
    \node[circle, fill=pink!75, inner sep=0pt, minimum size=0.12cm] (v2e) at (1,0.5) {};
    
    \node[circle, fill=pink!75, inner sep=0pt, minimum size=0.12cm] (v1w) at (-1,0) {};
    \node[circle, fill=pink!75, inner sep=0pt, minimum size=0.12cm] (v2w) at (-1,0.5) {};
    
    \node[circle, fill=pink!75, inner sep=0pt, minimum size=0.12cm] (v1n) at (-0.25,1.25) {};
    \node[circle, fill=pink!75, inner sep=0pt, minimum size=0.12cm] (v2n) at (0.25,1.25) {};
    
    \node[circle, fill=pink!75, inner sep=0pt, minimum size=0.12cm] (v1s) at (-0.25,-0.75) {};
    \node[circle, fill=pink!75, inner sep=0pt, minimum size=0.12cm] (v2s) at (0.25,-0.75) {};
    \draw[gray] (v1c) -- (v2e);
    \draw[gray] (v2c) -- (v1e);
    \node[] () at (0, -1.5) {\footnotesize Step $6$};
\end{tikzpicture}
$\quad$
\begin{tikzpicture}[scale=0.6]
    \draw[pink!75, fill=pink!40, rounded corners=2,thick]
     (-1.25, -0.25) rectangle (-0.75,0.75) {};
     \draw[pink!75, fill=pink!40, rounded corners=2,thick]
     (-0.5, 1) rectangle (0.5,1.5) {};
     \draw[pink!75, fill=pink!40, rounded corners=2,thick]
     (0.75, -0.25) rectangle (1.25,0.75) {};
     \draw[pink!75, fill=pink!40, rounded corners=2,thick]
     (-0.5, -1) rectangle (0.5,-0.5) {};
    \draw[celeste, fill=celeste!40, rounded corners=2,thick]
     (-0.25, -0.25) rectangle (0.25,0.75) {};
    \node[circle, fill=celeste, inner sep=0pt, minimum size=0.12cm] (v1c) at (0,0) {};
    \node[circle, fill=celeste, inner sep=0pt, minimum size=0.12cm] (v2c) at (0,0.5) {};
    
    \node[circle, fill=pink!75, inner sep=0pt, minimum size=0.12cm] (v1e) at (1,0) {};
    \node[circle, fill=pink!75, inner sep=0pt, minimum size=0.12cm] (v2e) at (1,0.5) {};
    
    \node[circle, fill=pink!75, inner sep=0pt, minimum size=0.12cm] (v1w) at (-1,0) {};
    \node[circle, fill=pink!75, inner sep=0pt, minimum size=0.12cm] (v2w) at (-1,0.5) {};
    
    \node[circle, fill=pink!75, inner sep=0pt, minimum size=0.12cm] (v1n) at (-0.25,1.25) {};
    \node[circle, fill=pink!75, inner sep=0pt, minimum size=0.12cm] (v2n) at (0.25,1.25) {};
    
    \node[circle, fill=pink!75, inner sep=0pt, minimum size=0.12cm] (v1s) at (-0.25,-0.75) {};
    \node[circle, fill=pink!75, inner sep=0pt, minimum size=0.12cm] (v2s) at (0.25,-0.75) {};
    \draw[gray] (v1c) -- (v2s);
    \draw[gray] (v2c) -- (v1s);
    \node[] () at (0, -1.5) {\footnotesize Step $7$};
\end{tikzpicture}
$\quad$
\begin{tikzpicture}[scale=0.6]
    \draw[pink!75, fill=pink!40, rounded corners=2,thick]
     (-1.25, -0.25) rectangle (-0.75,0.75) {};
     \draw[pink!75, fill=pink!40, rounded corners=2,thick]
     (-0.5, 1) rectangle (0.5,1.5) {};
     \draw[pink!75, fill=pink!40, rounded corners=2,thick]
     (0.75, -0.25) rectangle (1.25,0.75) {};
     \draw[pink!75, fill=pink!40, rounded corners=2,thick]
     (-0.5, -1) rectangle (0.5,-0.5) {};
    \draw[celeste, fill=celeste!40, rounded corners=2,thick]
     (-0.25, -0.25) rectangle (0.25,0.75) {};
    \node[circle, fill=celeste, inner sep=0pt, minimum size=0.12cm] (v1c) at (0,0) {};
    \node[circle, fill=pink!75, inner sep=0pt, minimum size=0.12cm] (v2n) at (0.25,1.25) {};
    \draw[gray] (v1c) -- (v2n);
    \draw[gray] (v2c) -- (v1n);
    
    \node[circle, fill=celeste, inner sep=0pt, minimum size=0.12cm] (v2c) at (0,0.5) {};
    \node[circle, fill=pink!75, inner sep=0pt, minimum size=0.12cm] (v1n) at (-0.25,1.25) {};
    
    \node[circle, fill=pink!75, inner sep=0pt, minimum size=0.12cm] (v1e) at (1,0) {};
    \node[circle, fill=pink!75, inner sep=0pt, minimum size=0.12cm] (v2e) at (1,0.5) {};
    
    \node[circle, fill=pink!75, inner sep=0pt, minimum size=0.12cm] (v1w) at (-1,0) {};
    \node[circle, fill=pink!75, inner sep=0pt, minimum size=0.12cm] (v2w) at (-1,0.5) {};

    \node[circle, fill=pink!75, inner sep=0pt, minimum size=0.12cm] (v1s) at (-0.25,-0.75) {};
    \node[circle, fill=pink!75, inner sep=0pt, minimum size=0.12cm] (v2s) at (0.25,-0.75) {};
    \node[] () at (0, -1.5) {\footnotesize Step $8$};
\end{tikzpicture}
    
    \caption{Gate sequence of the $\llbracket2,1,1\rrbracket$ scheme.}
\label{fig:sequence}
\end{figure}
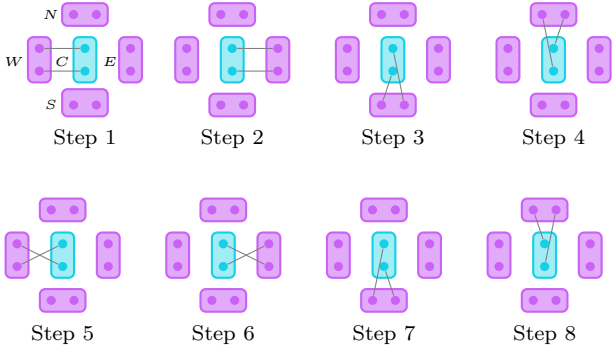

\renewcommand{\arraystretch}{1.5}
\begin{table}[h]
\begin{center}
\begin{tabularx}{\columnwidth}{|X          | X         |X          |X        |} 
\hline
Inner codes     &Cubic      &$\llbracket4,1,1,2\rrbracket$&$\llbracket7,1,3\rrbracket$          \\
\hline
Stabilizers           &           &  $XXXX$   &  $g_1,\cdots,g_6$   \\
                &           &  $ZZZZ$   &                     \\
\hline
$\overline{X}$  &$X$        &  $XXII$   &  $X_1\cdots X_7$   \\
$\overline{Z}$  &$Z$        &  $ZIZI$   &  $Z_1\cdots Z_7$   \\    
\hline
\end{tabularx}
\end{center}

\begin{center}
\begin{tabularx}{\columnwidth}{|X          | X        |X          |X        |} 
\hline
 Inner codes    &$\llbracket2,1,1\rrbracket$&$\llbracket3,1,1\rrbracket_1$&$\llbracket3,1,1\rrbracket_2$      \\
\hline
Stabilizers     &  $XX$     & $XXI$       & $XXX$             \\
                &           & $IXX$       & $IZZ$             \\
\hline
$\overline{X}$  &  $XI$     & $XII$       & $XII$             \\
$\overline{Z}$  &  $ZZ$     & $ZZZ$       & $ZIZ$             \\    
\hline
\end{tabularx}
\caption{Definitions of the inner codes in the stabilizer formalism and their respective logical operators.}
\label{tab:stabs}
\end{center}
\end{table}

\section{Thresholds}
\label{sec:sim}
In this Section, we discuss the performance of our schemes against the phenomenological noise model and the circuit-level noise model in terms of the threshold error rate. After measuring all the qubits in the $X$-basis, the decoder works in two steps. First, if an error is detected in any individual code block, that code block is erased. If not, the measured logical information of that code block is passed to the higher level decoder, which we choose to be the MWPM decoder with Manhattan norm. We use the Blossom V algorithm~\cite{kolmogorov2009blossom} with $10^6$ trials to get each data point. We perform the simulation on lattices of varying size $L$, ranging from $3$ to $19$.

\subsection{Phenomenological noise model}
\label{conversion relations}

The phenomenological noise model is a model in which the noiseless cluster state $|\psi_{G,\mathcal{C}}\rangle$ is subjected to a depolarizing noise channel $\mathcal{D}_p$:
\begin{equation}
    \mathcal{D}_p(\rho) = (1-\frac{3}{2}p)\rho + \frac{p}{2}(X\rho X + Y\rho Y + Z\rho Z)
    \label{equ:single_qubit_depol}
\end{equation}
on each qubit\footnote{We choose the normalization such that the value of $p$ corresponds to the probability of an $X$ error occurring. It's important to note that for our purposes, we treat a $Y$ error as being equivalent to an $X$ error followed by a $Z$ error.}. We expect the logical error rate $p_L$ to obey the following scaling relation near the threshold:
\begin{equation}
    p_{L}=f(a+b(p-p_{th})d^\frac{1}{\nu}).
\end{equation}
Fitting to the quadratic scaling ansatz $\sum_{i=0}^{3} A_i (p_i-p_{th})d^\frac{1}{\nu}$~\cite{Wang_2003} with parameters $A_i$, $p_{th}$, and $\nu$, we obtain a threshold of $8.034(3)\%$.

\begin{figure}[h]
\centering\includegraphics[scale=0.025]{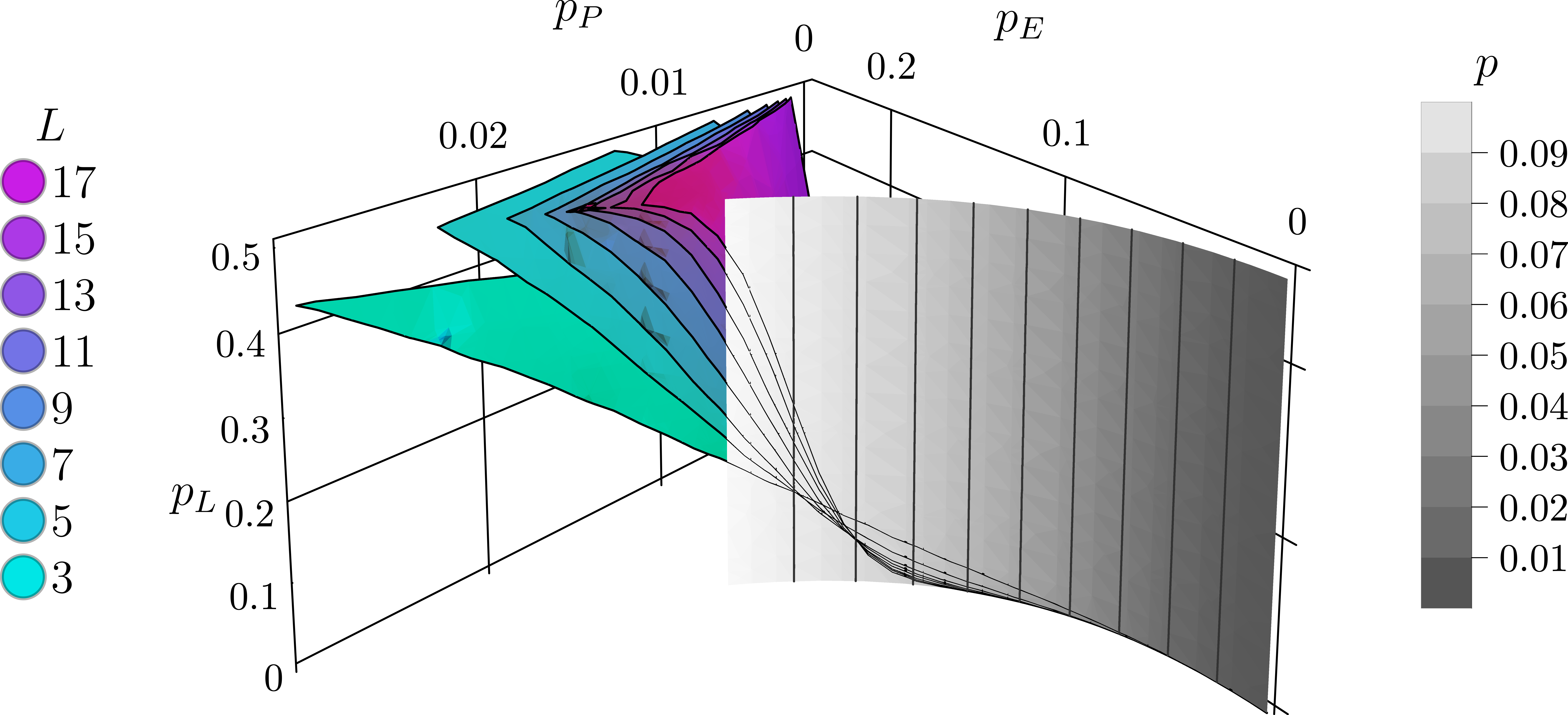}
\caption{Phase diagram for the bcc scheme~\cite{raussendorf2006fault,raussendorf2007fault,raussendorf2007topological}, here $p_E$ and $p_P$ are the loss and Pauli error rates, respectively, and $p_L$ is the logical error rate. The grayscale surface is the section of effective loss-error relationship for the $\llbracket2,1,1\rrbracket$ scheme, from which we can directly read off the error threshold of approximately $8.03\%$.}
\label{fig:plot3D}
\end{figure}

Intuitively, we can understand the value of this threshold in terms of an effective error model on the bcc cluster state. For instance, for the $\llbracket2,1,1\rrbracket$ scheme, upon converting the error detected from the code $\mathcal{C}$ to an erasure error, we obtain an effective model in which the rates for Pauli ($p_P$) and erasure ($p_E$) errors are
\begin{equation}
\begin{aligned}
    p_P &= \frac{p^2}{1-2p(1-p)}, \\
    p_E &= 2p(1-p).
\end{aligned}
\end{equation}
The model effectively defines a path along the phase diagram of the bcc cluster state as shown in Figure~\ref{fig:plot3D}. We note that the values of the effective $p_P$ and $p_E$ obtained at our threshold are consistent with the thresholds for the respective parameters obtained in Ref.~\cite{barrett2010fault}. The conversion of the Pauli errors into loss errors generally leads to improvements in thresholds, as summarized in Table~\ref{tab:indep}.

\renewcommand{\arraystretch}{1.5}
\begin{table}[h]
\centering\begin{tabularx}{\columnwidth}{|X          | X            |X          |X       |} 
\hline
Inner codes     &Cubic       &$\llbracket4,1,1,2\rrbracket$  &$\llbracket7,1,3\rrbracket$          \\
\hline
$p_{th}$         &$2.936(2)\%$\footnote{Our data agrees with the reported results from \cite{Wang_2003}}&$4.195(5)\%$  & $4.137(5)\%$          \\
\hline
\end{tabularx}
\centering\begin{tabularx}{\columnwidth}{|X          |X            |X          |X        |} 
\hline
Inner codes    &$\llbracket2,1,1\rrbracket$ &$\llbracket3,1,1\rrbracket_1$&$\llbracket3,1,1\rrbracket_2$        \\
\hline
$p_{th}$         &$8.034(3)\%$ &$10.26(1)\%$&$5.66(1)\%$          \\
\hline
\end{tabularx}
\caption{Thresholds of the concatenation schemes against phenomenological noise.}
\label{tab:indep}
\end{table}

\subsection{Circuit-level noise model}
\label{sec:circuit_level}
A more realistic error model should take into account errors in individual operations, which we analyze as follows. We assume that the logical $|\overline{+}\rangle$ state of the code $\mathcal{C}$ is prepared perfectly, followed by independent depolarizing noise on every physical qubit. After that, every gate is followed by depolarizing noise, i.e. each two-qubit error can occur with probability $p/15$, and measurements are preceded by depolarizing noise as well. Finally, we also assume that idle qubits experience depolarizing noise.\footnote{We note in passing that, unlike the other codes, the $\llbracket3,1,1\rrbracket_2$ code requires one more idle timestep for two of the qubits, in order to have consistent global scheduling of the gates. We have ignored the extra error incurred in this step. However, the conclusion of our work does not depend significantly on this choice.}

One may wonder about the motivation behind our non-standard choice of modeling the logical $|\overline{+}\rangle$ state preparation as a perfect preparation followed by depolarizing noise. This is because there are known methods to boost the fidelity of such states using methods such as post-selection~\cite{reichardt2006postselection,chao2018quantum,ryan2021realization}. As such, the result of these simulations can be thought of as the best-case-scenario threshold one can hope to get, after incorporating such techniques.

We also note that such an error model reduces to the standard circuit-level depolarizing noise model if the logical $|\overline{+}\rangle$ state is a product state. In our case, there are two examples: the $\llbracket2,1,1\rrbracket$ and the $\llbracket3,1,1\rrbracket_1$ code. Therefore, for these codes, one can interpret the result as the true threshold against the standard circuit-level noise model.

After the state is prepared, all the qubits are measured in the $X$-basis. Applying the decoding algorithm in Section~\ref{sec:concatenation}, we estimate the logical error rate; see Table~\ref{tab:circ} for the summary. We note that the thresholds are generally better than the bcc cluster state, except for the $\llbracket3,1,1\rrbracket_1$-code-based scheme. 

Taking into account our non-standard logical $|\overline{+}\rangle$ state preparation error modeling, we can interpret these results as follows. First of all, because our model reduces to the standard circuit-level noise model for the $\llbracket2,1,1\rrbracket$ and the $\llbracket3,1,1\rrbracket_1$ code, these results can be compared fairly against the other schemes in the literature. We note that the $\llbracket2,1,1\rrbracket$ scheme improves upon the threshold of the bcc scheme, whereas the $\llbracket3,1,1\rrbracket_1$ scheme does not. Since repetition codes involving a larger number of qubits will require even more gates, we can therefore expect that $\llbracket n,1,1\rrbracket$ code-based constructions for $n>3$ will likely yield worse results. 

\begin{figure}[h]
\begin{subfigure}{\columnwidth}
\centering\includegraphics[scale=0.8]{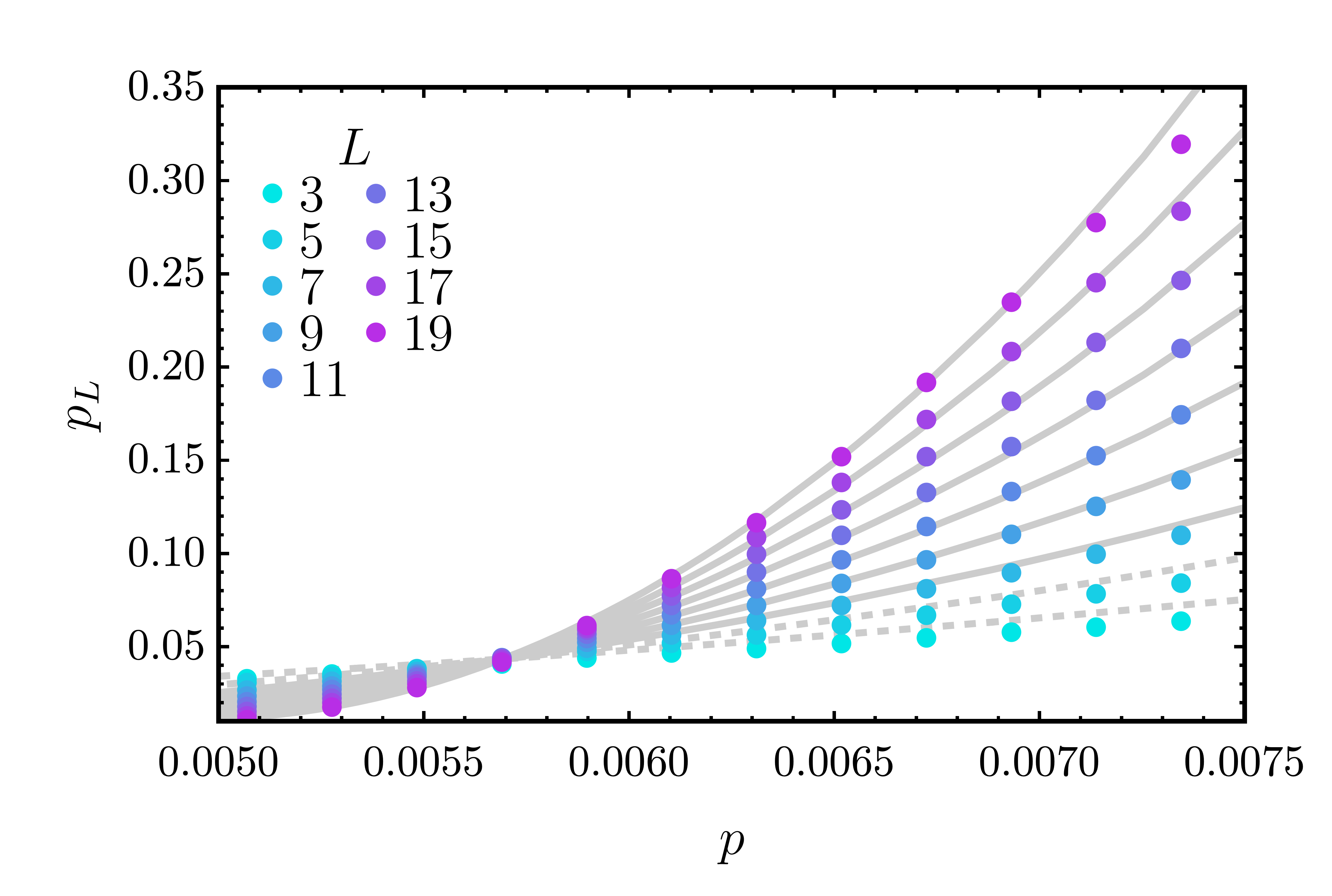}
\caption{Error threshold of the bcc scheme}
\end{subfigure}
\begin{subfigure}{\columnwidth}
\centering\includegraphics[scale=0.8]{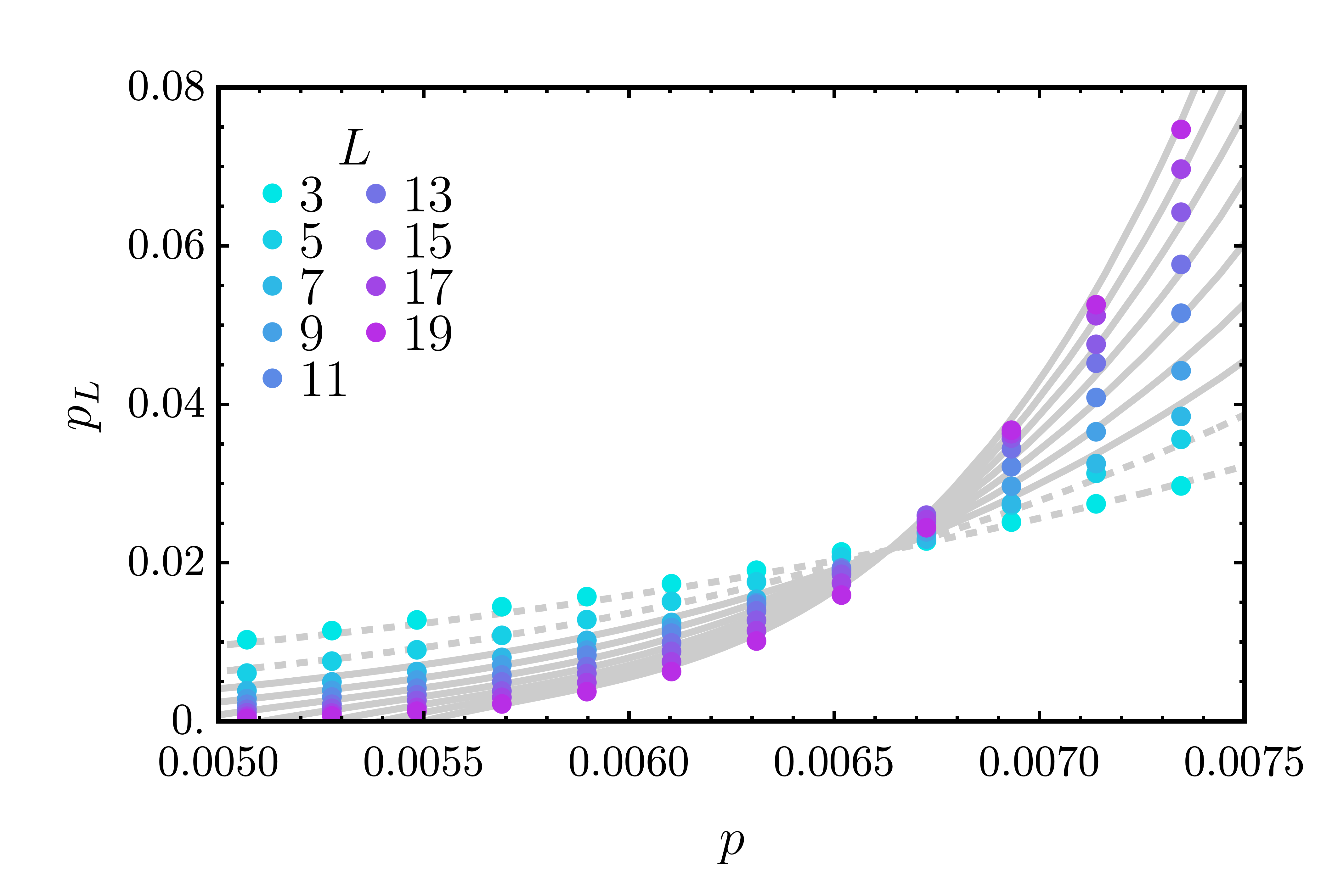}
\caption{Error threshold of the $\llbracket2,1,1\rrbracket$ scheme}
\end{subfigure}
\caption{Comparison of error threshold under circuit-level noise\label{fig:thresholdcomp}. Note that the dashed lines in the figure represent data that suffer from finite size effects and are therefore excluded from the fitting process.}
\end{figure}

Secondly, for the other schemes, e.g., $\llbracket3,1,1\rrbracket_2,$ $\llbracket4,1,1,2\rrbracket$, and $\llbracket7,1,3\rrbracket$, we see improvements in the threshold. Therefore, if we can utilize the high-fidelity logical state preparation schemes, e.g., Ref.~\cite{reichardt2020fault,ryan2022implementing},
we can expect to again have improvements against the bcc scheme. However, we note that a more naive state preparation circuit does not yield improvements. For instance, we studied the state preparation circuit for the $\llbracket4,1,1,2\rrbracket$ in which one prepares the logical $|\overline{+}\rangle |\overline{0}\rangle$ state, which can be prepared fault-tolerantly by preparing two Bell pairs. Even for such a simple circuit, we observed a lower threshold than the bcc cluster state, which is shown in Figure~\ref{fig:thresholdcomp}.

\renewcommand{\arraystretch}{1.5}
\begin{table}[h]
\begin{center}
\begin{tabularx}{\columnwidth}{|X          | X            |X          |X        |} 
\hline
Inner codes     &Cubic       &$\llbracket4,1,1,2\rrbracket$  &$\llbracket7,1,3\rrbracket$          \\
\hline
$p_{th}$        &$0.5692(1)\%$&$0.701(1)\%$ & $0.678(1)\%$        \\
\hline
\end{tabularx}
\begin{tabularx}{\columnwidth}{|X          | X            |X          |X        |} 
\hline
Inner codes    &$\llbracket2,1,1\rrbracket$  &$\llbracket3,1,1\rrbracket_1$&$\llbracket3,1,1\rrbracket_2$        \\
\hline
$p_{th}$        &$0.664(3)\%$ &$0.3216(1)\%$&$0.6947(2)\%$          \\
\hline
\end{tabularx}
\end{center}
\caption{Threshold of the concatenation schemes against circuit-level noise model, assuming that the initial encoded logical state of the code is prepared perfectly, followed by a depolarizing noise model on all the qubits.}
\label{tab:circ}
\end{table}

\subsection{Biased noise model}

In this section, we study our schemes against the biased noise. We remark that the following analysis is not based on an explicit Monte Carlo simulation. Rather, it is based on an exact mapping of the biased noise model to the phenomenological noise model by exploiting the fact that $Z$ errors do not propagate through the circuit. 

In this model, the two-qubit noise in Section~\ref{sec:circuit_level} is replaced with an error model completely biased to $Z$-error. For instance, for the two-qubit gates, we apply the following noise model. 
\begin{align}
    \mathcal{D}_p^Z(\rho)=(1-p)\rho+\frac{p}{3}(ZZ\rho ZZ+ZI\rho ZI+IZ\rho IZ)
\end{align} 
This error model is equivalent to a limiting case of the error model studied in Ref.~\cite{stephens2013high}, with the bias parameter $\beta$ taken to be infinity, i.e. we essentially suppress the propagation of gate level noise. Subsequently, the model reduces to an independent error model whose resulting threshold error rate can be calculated by a counting argument. For instance, the $\llbracket2,1,1\rrbracket$ scheme results in an effective noise parameter of $p=4\cdot2\cdot p_{CZ}+p_M+p_I=4\cdot2\cdot2p/3+4p/3=20p/3$, leading to a $p_{th}$ of $1.205(1)\%$~\footnote{Our results point out a possible systematic inconsistency in Ref.~\cite{stephens2013high}, as the presented numerical results exceed the worst-case scenario delineated by our semi-analytic approach.}. Unsurprisingly, the scheme is extremely tolerant to biased noise, as indicated by the effective error rate listed in Table~\ref{tab:biased}. 

\renewcommand{\arraystretch}{1.5}
\begin{table}[h]
\begin{center}
\begin{tabularx}{\columnwidth}{|X          | X            |X          |X        |} 
\hline
Inner codes     &Cubic       &$\llbracket4,1,1,2\rrbracket$  &$\llbracket7,1,3\rrbracket$          \\
\hline
Biased          &$0.734(5)\%$&$1.05(1)\%$  & $1.03(2)\%$        \\
\hline
\end{tabularx}

\begin{tabularx}{\columnwidth}{|X          | X            |X          |X        |} 
\hline
Inner codes    &$\llbracket2,1,1\rrbracket$ &$\llbracket3,1,1\rrbracket_1$&$\llbracket3,1,1\rrbracket_2$        \\
\hline
Biased          &$1.205(1)\%$ &$1.090(5)\%$&$1.18(5)\%$          \\
\hline
\end{tabularx}
\end{center}
\caption{Thresholds of the concatenation schemes against biased noise model.}
\label{tab:biased}
\end{table}

\section{Overhead}
\label{sec:analytics}

We now study the overhead of our schemes assuming a range of aspirational but reasonable circuit-level noise rates. Overall, we note that the only scheme that leads to a lower overhead is based on the $\llbracket2,1,1\rrbracket$ code. As such, we primarily focus on this code.

A particularly interesting regime is an error rate approaching $10^{-3}$. This error rate is slightly lower than what can currently be achieved experimentally. As such, studies of overhead in this parameter range can provide an accurate estimate of the overhead required in realistic quantum computers. 

Normally, the logical error rate decays exponentially with the code distance $L$. However, our numerical study shows that there are effects that make the fitting to an exponentially decaying function unreliable, which may be a result of the interplay between the decoder and the inner code~\cite{fowler2012topological}. Thus, we first focus on demonstrating improvements in overheads in the high-error-rate regime, based on our Monte Carlo simulation. We then explain, based on a combinatorial argument, why we anticipate this improvement to persist in a lower error rate regime as well.

\begin{figure}[H]
\includegraphics[scale=0.8]{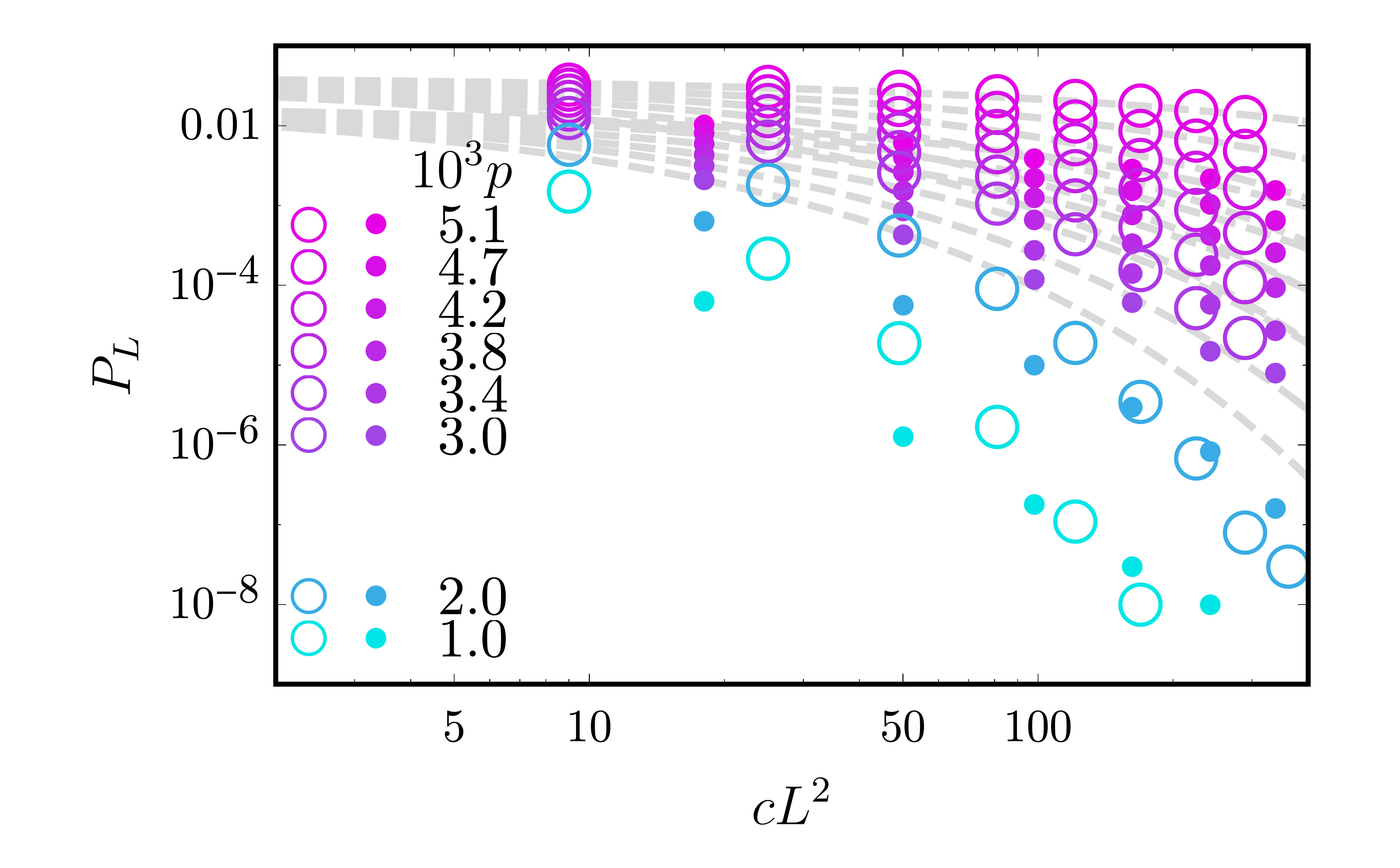}
\caption{Relationship of $p_L$ v.s. $L$ at different $p$ values. Here the dots represent data points of the $\llbracket2,1,1\rrbracket$ scheme and the circles represent that of the bcc scheme.}
\label{fig:overhead}
\end{figure}

Due to the aforementioned effects, we cannot reliably extrapolate the suppression of $p_L$, and the overhead advantage may not exist for large $L$. As shown in Figure~\ref{fig:overhead}, the error suppresion relation for the $\llbracket2,1,1\rrbracket$ scheme are convex in the log-log plot and eventually crosses that of the bcc scheme. However, we can instead interpolate the data to fairly compare the overhead of our $\llbracket2,1,1\rrbracket$ scheme to the bcc cluster state when $L$ is not that large. We fit the logical error rate at every physical error rate with the function $e^{a+bL+cL^3}$ to account for those effects~\footnote{The choice of the qubic non-linear term is to accommodate to the numerical methods.}. From this, we can infer overhead for the overall spacetime volume scaling as $sL^2$, where $s$ is the block size of the inner code. The $\llbracket2,1,1\rrbracket$ scheme outperforms the bcc scheme in this regime, as shown in Figure~\ref{fig:overhead}.

For smaller and more practical $p_{th}$ values, numerical simulation fails to give accurate results due to the diverging time to convergence. However, the following analytical argument, based on the leading-order approximation, suggests that even in this regime we expect to have lower overhead.

For the bcc lattice with periodic boundary conditions in two directions and a smooth boundary on the third, $p_L$ for a single logical $X$ operator, to the leading order, scales approximately as~\cite{watson2014logical}: 
\begin{align}
\label{3D analytic}
    L^2\left(\begin{aligned}
    &\,L\\
   \biggl\lfloor&\frac{L}{2}\biggl\rfloor
\end{aligned}\right)p^{\floor*{L/2}}\sim \frac{L^{3/2}}{\sqrt{2\pi}}(p/4)^{\floor*{L/2}},
\end{align} where $L^2$ counts the number of lowest weight logical operators that percolate to the two rough boundaries. Thus $p_L=o(p^{L/2})$ as $p\rightarrow 0$ and $L\rightarrow\infty$. 

For the $\llbracket2,1,1\rrbracket$ scheme with the same boundary condition, a length $L$ path could fit in $N\in\{1,\cdots,L\}$ erasure errors, and the effective code distance is reduced to $L-N$. Using the effective $p_P$ and $p_L$ relations outlined in Subsection \ref{conversion relations}, $p_L$ is given by 
\begin{align}
   L^2\sum_{N=0}^L\left(\begin{aligned}
    L\\
    N
\end{aligned}\right)\left(\begin{aligned}
    &\,L-N\\
    \biggl\lfloor&\frac{L-N}{2}\biggr\rfloor
\end{aligned}\right)2^Np^{N+2\floor*{\frac{L-N}{2}}},
\end{align} which scales as $o(p^L)$ as $p\rightarrow 0$ and $L\rightarrow\infty$. Moreover, this quantity is upper-bounded by 
\begin{align}
    L^3\left(\begin{aligned}
    L\\
\frac{L}{2}
\end{aligned}\right)^2p^L\sim \frac{L^2}{2\pi}(p/4)^{L}
\end{align}
as $L\rightarrow\infty$, which is, in turn, upper-bounded by Eq. \ref{3D analytic}. 

Since the $\llbracket2,1,1\rrbracket$ scheme only contributes to a constant factor in the overhead, in the low-noise regime we expect to have better performance in terms of resource usage. For the other concatenated schemes, due to $\mathcal{C}$-detectability, the logical Pauli error rate $p$ scales as $o(p^i)$ with some $i\geq2$. Therefore, we expect similar results to hold. At around $p_L=10^{-6}$ and $p=10^{-3}$, the overhead value is listed in Table~\ref{tab:overhead}. 

\renewcommand{\arraystretch}{1.5}
\begin{table}[h]
\begin{center}
\begin{tabularx}{\columnwidth}{|X          |X            |X          |X        |} 
\hline
Inner codes     &Cubic       &$\llbracket4,1,1,2\rrbracket$  &$\llbracket7,1,3\rrbracket$          \\
\hline
$L_c/L$ indep.    &$1.00$   &$1.27$   & $2.40$       \\
\hline
\end{tabularx}
\begin{tabularx}{\columnwidth}{|X          | X            |X          |X       |} 
\hline
Inner codes    &$\llbracket2,1,1\rrbracket$ &$\llbracket3,1,1\rrbracket_1$&$\llbracket3,1,1\rrbracket_2$        \\
\hline
$L_c/L$ indep.    &$0.68$   &$4.74$    & $1.43$        \\
\hline
\end{tabularx}
\end{center}
\caption{Ratio of overhead between concatenation schemes $L_c$ and the bcc scheme $L$.}
\label{tab:overhead}
\end{table}

\section{Conclusions}
We have identified a fault-tolerant quantum error correction scheme that effectively converts circuit-level Pauli errors to erasure errors. Besides the much higher phenomenological error threshold compared with regular foliated topological quantum codes~\cite{bolt2016foliated}, our scheme also demonstrates improvements in circuit-level error thresholds, indicating a path to practical performance advantage. We have restricted our discussion to decoding by the MWPM algorithm with Manhattan norm weightings. Using better decoding schemes, employing lattice correlations, or utilizing predecoders will likely lead to even more improvements~\cite{breuckmann2018scalable,wang2011surface,chamberland2022techniques,smith2022local,das2022afs,delfosse2020hierarchical}.
Furthermore, different outer codes, such as those with $>50\%$ erasure thresholds~\cite{nickerson2018measurement}, could be used in similar constructions\footnote{Even a trivial cluster state consisting of $n$ EPR pairs has a threshold of 100$\%$. This does not violate the no-cloning theorem since no logical state is encoded in the lattice, but only a logical EPR pair. In fact, the 2-way quantum capacity of a binary erasure channel is exactly 1.}. Lastly, we leave it as an open problem to convert our scheme to a two-dimensional architecture and study its performance against the standard surface code.

\section{Acknowledgement}
We would like to express our gratitude to Kianna Wan, Noah Shutty, Scott Xu, Ani Krishna, and Michael Vasmer for their valuable insights and helpful discussions. We also thank the anonymous referees for their detailed feedback and constructive suggestions that have greatly improved the quality of this manuscript. P.H. acknowledges the support of ARO (award W911NF2120214), CIFAR, and the Simons Foundation.
\bibliography{main.bib}
\onecolumngrid
\newpage
\appendix
\section{Effective error model}
The effective circuit-level error model is obtained by conjugating all single-qubit and two-qubit error operators to the end of the circuit. Here, the tables show the errors defined by the noise model and the resulting effective error after propagation.
\label{appendix:a}
\renewcommand{\arraystretch}{1}
\begin{table}[h]
\begin{minipage}[t]{.4\textwidth} %
\begin{tabularx}{\textwidth}{ |c|c|b| }
\hline
Error & Step & Effective error        \\
\hline
$X_C$& &$Z_W$  			 		   \\
$X_CZ_W$&1&       			 	   \\
$Z_W$& & $Z_W$           		        \\ 
\hline    
$X_C$& &$Z_W$      		 		   \\
$X_CZ_E$&2&$Z_W$,$Z_E$      	   \\
$Z_E$& &$Z_E$     	     		   \\
\hline
$X_C$& &$Z_W$,$Z_E$      		    \\
$X_CZ_N$&3&$Z_W$,$Z_E$,$Z_N$	   \\
$Z_N$& &$Z_N$     		 		   \\
\hline
$X_C$& &$Z_W$,$Z_E$,$Z_N$		   \\
$X_CZ_S$&4&$Z_W$,$Z_E$,$Z_N$,$Z_S$\\
$Z_S$& &$Z_S$     		          \\
\hline
\end{tabularx}
\caption{Propagated circuit-level errors for type I codes with the trivial circuit.}
\end{minipage}%
\hspace{.04\textwidth}
\begin{minipage}[t]{.56\textwidth}
\begin{tabularx}{\textwidth}{ |c|c|b||c|b|  }
\hline
 Error & Step  & Effective error &Step & Effective error\\
\hline
$X_C$   & &$Z_W$   		  		    &   &  $Z_E$,$Z_N$,$Z_S$     \\
$X_CZ_W$&1&    		  		        &  5&  $Z_W$,$Z_E$,$Z_N$,$Z_S$  	  \\
$Z_W$&  &  $Z_W$ 	  		  		&   &  $Z_W$      			  \\ 
\hline   	
$X_C$& &$Z_W$      		  	        &   &   $Z_N$,$Z_S$	        \\
$X_CZ_E$&2&$Z_W$,$Z_E$        	    &  6&  $Z_E$,$Z_N$,$Z_S$  \\
$Z_E$& &$Z_E$      		   	        &   &  $Z_E$      			  \\
\hline
$X_C$& &$Z_W$,$Z_E$ 	  	        &   &   $Z_S$  		  \\
$X_CZ_N$&3&$Z_W$,$Z_E$,$Z_N$  	    &  7&  $Z_N$,$Z_S$      			  \\
$Z_N$& &$Z_N$      		  	        &   &  $Z_N$     			  \\
\hline
$X_C$& &$Z_W$,$Z_E$,$Z_N$  	        &   &  $Z_S$   			  	  \\
$X_CZ_S$&4&$Z_W$,$Z_E$,$Z_N$,$Z_S$  &  8&       			    \\
$Z_S$& &$Z_S$     			        &   & $Z_S$        \\
\hline
\end{tabularx}
\caption{Propagated circuit-level errors for the $\llbracket2,1,1\rrbracket$ scheme. The circuit employs the $\llbracket2,1,1\rrbracket$-detectable gate sequence demonstrated in the Fig.~\ref{fig:sequence}.}
\end{minipage}
\end{table}
\begin{table}[h]
\noindent\begin{tabularx}{\textwidth}{ |c|c|X||c|X||c|X|  } %
\hline
Error & Step & Effective error & Step & Effective error & Step & Effective error \\
\hline
$X_C$& &$Z_W$     				 & &   $Z_W$,$Z_E$,$Z_N$,$Z_S$          & &  $Z_E$,$Z_N$,$Z_S$         \\
$X_CZ_W$&1&\quad         		 &5&  $Z_{W12}$,$Z_E$,$Z_N$,$Z_S$        &9&   $Z_W$,$Z_E$,$Z_N$,$Z_S$  \\
$Z_W$& &$Z_W$                    & &   $Z_W$                             & &  $Z_W$                     \\
\hline   
$X_C$& &$Z_W$     				 & &  $Z_{W12}$,$Z_E$,$Z_N$,$Z_S$        & &    $Z_S$,$Z_N$             \\
$X_CZ_E$&2&$Z_W$,$Z_E$			 &6&  $Z_{W12}$,$Z_{E12}$,$Z_N$,$Z_S$    &10&  $Z_E$,$Z_N$,$Z_S$        \\
$Z_E$& &$Z_E$     				 & &  $Z_E$             			     & &  $Z_E$                     \\
\hline
$X_C$& &$Z_W$,$Z_E$    	   		 & &  $Z_W$,$Z_E$,$Z_N$,$Z_{S12}$        & &   $Z_S$   		            \\
$X_CZ_N$&3&$Z_W$,$Z_E$,$Z_N$     &7&  $Z_W$,$Z_E$,$Z_{N12}$,$Z_{S12}$    &11&  $Z_S$,$Z_N$       	    \\
$Z_N$& &$Z_N$      		  		 & &  $Z_N$                              & &  $Z_N$    	                \\
\hline
$X_C$& &$Z_W$,$Z_E$,$Z_N$  		 & &     $Z_W$,$Z_E$,$Z_N$,$Z_S$         & &     $Z_S$        	        \\
$X_CZ_S$&4&$Z_W$,$Z_E$,$Z_N$,$Z_S$&8&  $Z_W$,$Z_E$,$Z_N$,$Z_{S12}$	     &12&   \quad       		  	\\
$Z_S$& &$Z_S$     				 & &   $Z_S$                             &      &$Z_S$                  \\
\hline
\end{tabularx}
\caption{Circuit-level errors for the $\llbracket3,1,1\rrbracket_1$ scheme. The circuit is $\llbracket3,1,1\rrbracket_1$-detectable by applying transversal gate sequences for 3 rounds.}
\end{table}

\section{Scaling exponents}
The scaling exponent $\nu$ obtained from fitting is listed as follows.
\renewcommand{\arraystretch}{1.5}
\begin{table}[h]
\begin{center}
\begin{tabularx}{0.7\columnwidth}{|X          | X            |X          |X        | X            |X         |X        |} 
\hline
Inner codes     &Cubic       &$\llbracket4,1,1,2\rrbracket$  &$\llbracket7,1,3\rrbracket$   &$\llbracket2,1,1\rrbracket$ &$\llbracket3,1,1\rrbracket_1$&$\llbracket3,1,1\rrbracket_2$       \\
\hline
$\nu$ indep.    &$0.92(1)$   &$1.28(6)$    & $1.13(7)$    &$1.40(9)$   &$1.04(3)$    & $1.21(3)$    \\
\hline
$\nu$ circ.     &$0.88(1)$   &$1.26(3)$    & $1.06(2)$     &$1.29(7)$   &$1.04(2)$    &  $1.04(3)$   \\
\hline
\end{tabularx}
\end{center}
\caption{Scaling exponents of the concatenation schemes against the phenomenological and circuit-level noise model.}
\label{tab:exp}
\end{table}
\section{Threshold simulations}
The raw data obtained from our numerical simulation is plotted in the following diagrams. 
\begin{spacing}{0}
\begin{figure}[h]
\centering
    \begin{minipage}[t]{0.45\textwidth}
     \includegraphics[width=\linewidth]{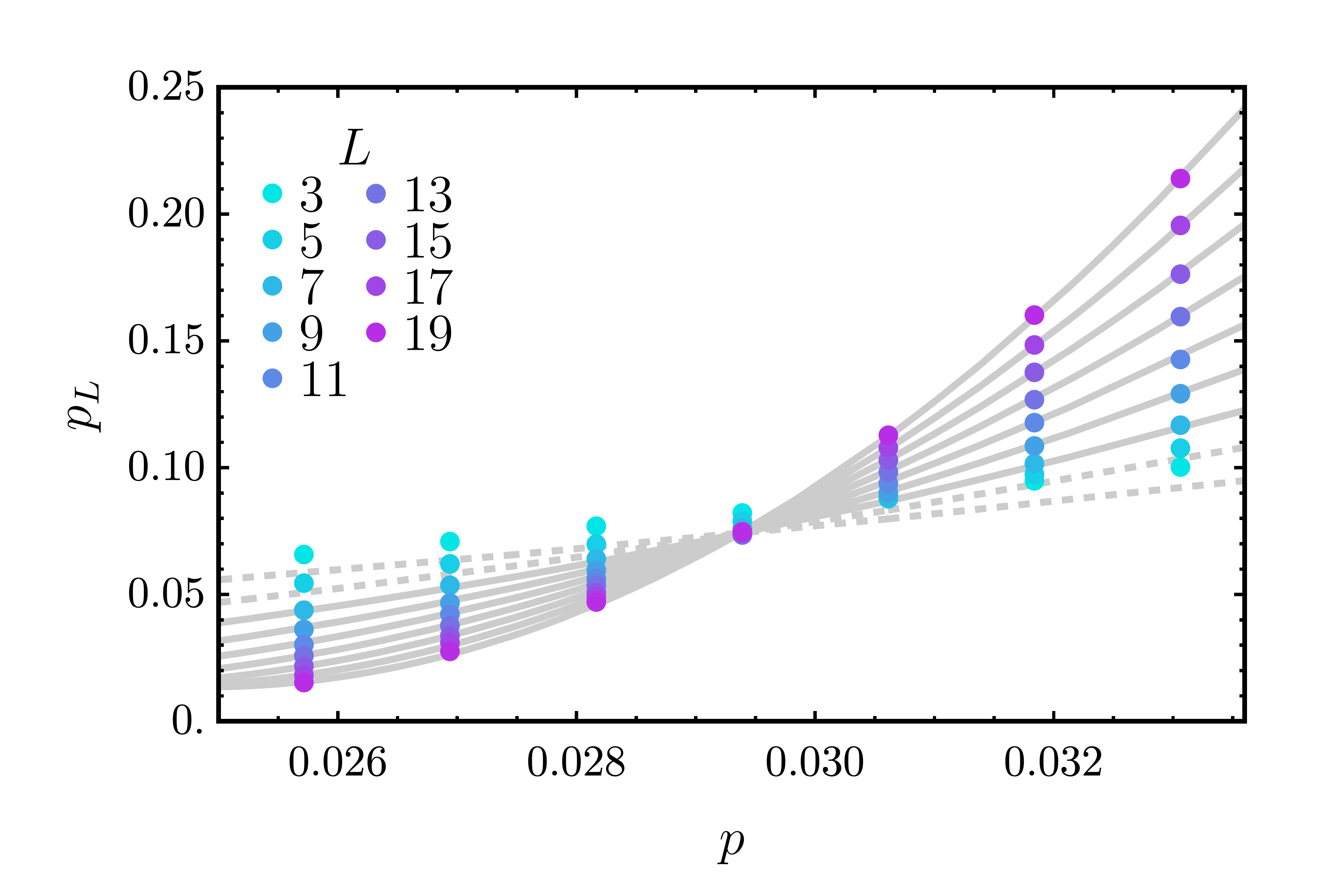}
     \caption{Logical error probability of the bcc scheme under the phenomenological noise model.}
    \end{minipage}\quad
    \begin{minipage}[t]{0.45\textwidth}
        \includegraphics[width=\linewidth]{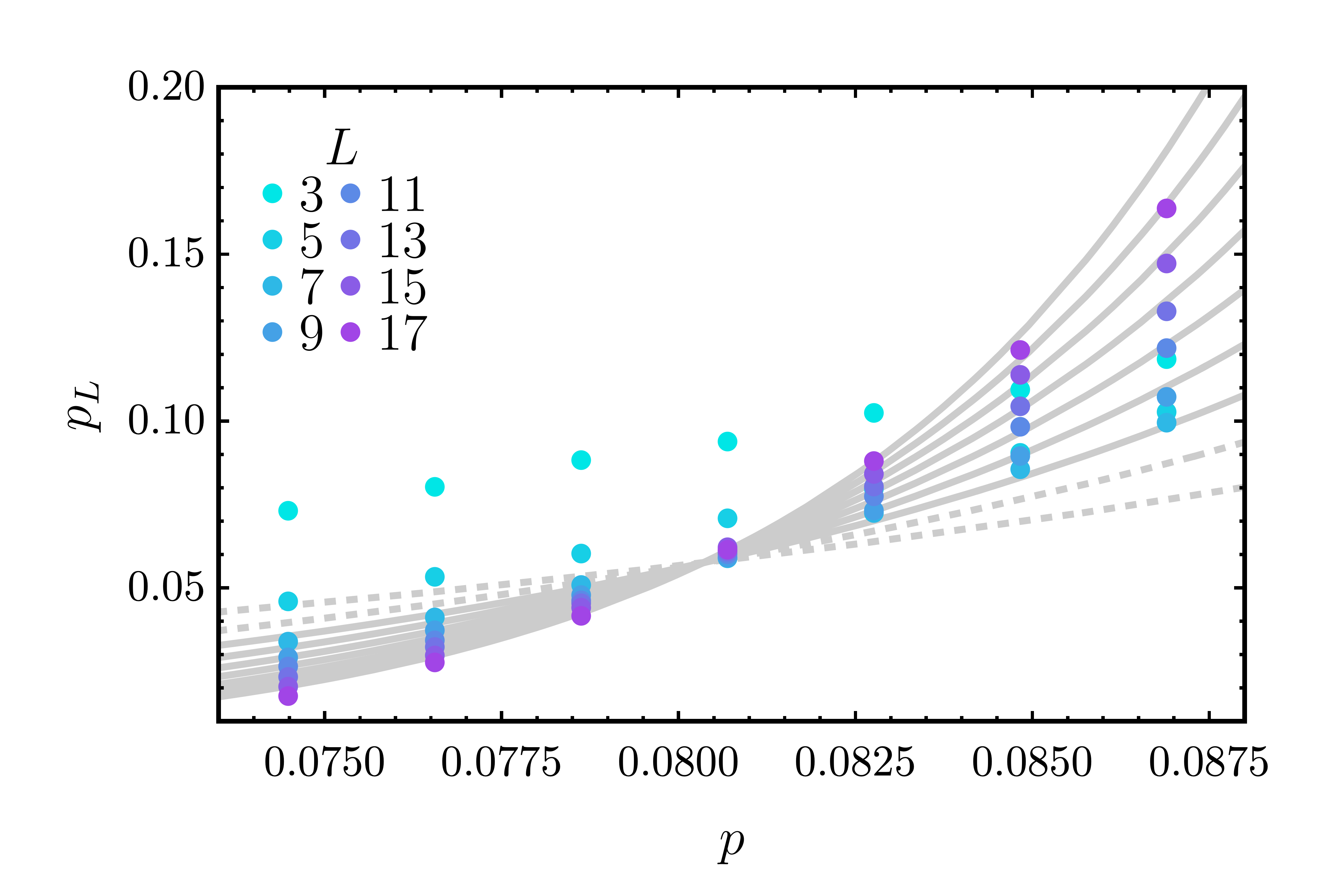}
        \caption{Logical error probability of the $\llbracket2,1,1\rrbracket$ scheme under the phenomenological noise model.}
    \end{minipage}\quad
    \label{fig:example}%
\end{figure}\quad

\begin{figure}[h]
\centering
    \begin{minipage}[t]{0.45\textwidth}
        \includegraphics[width=\linewidth]{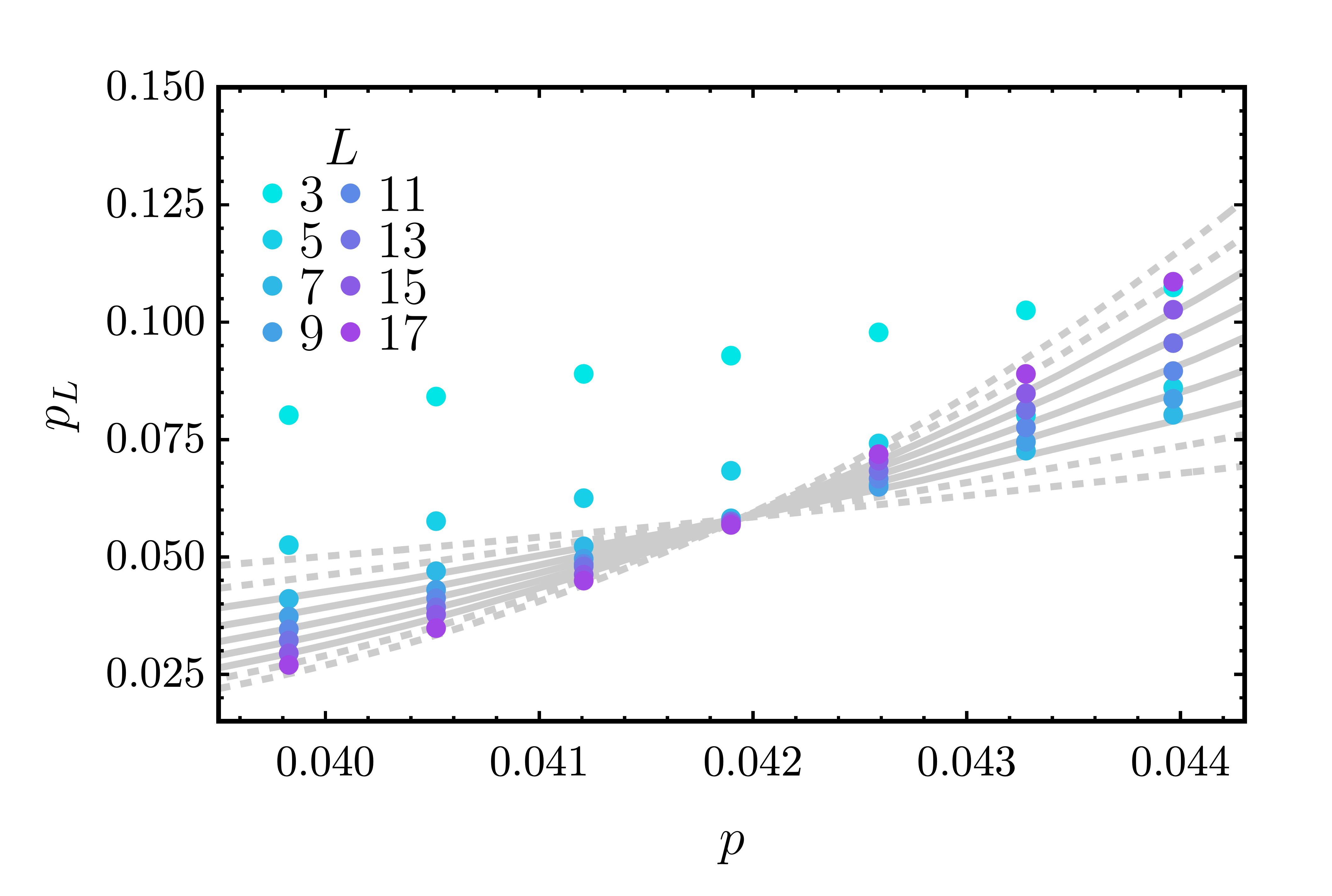}
        \caption{Logical error probability of the $\llbracket4,1,1,2\rrbracket$ scheme under the phenomenological noise model.}
    \end{minipage}\quad
    \begin{minipage}[t]{0.45\textwidth}
        \includegraphics[width=\linewidth]{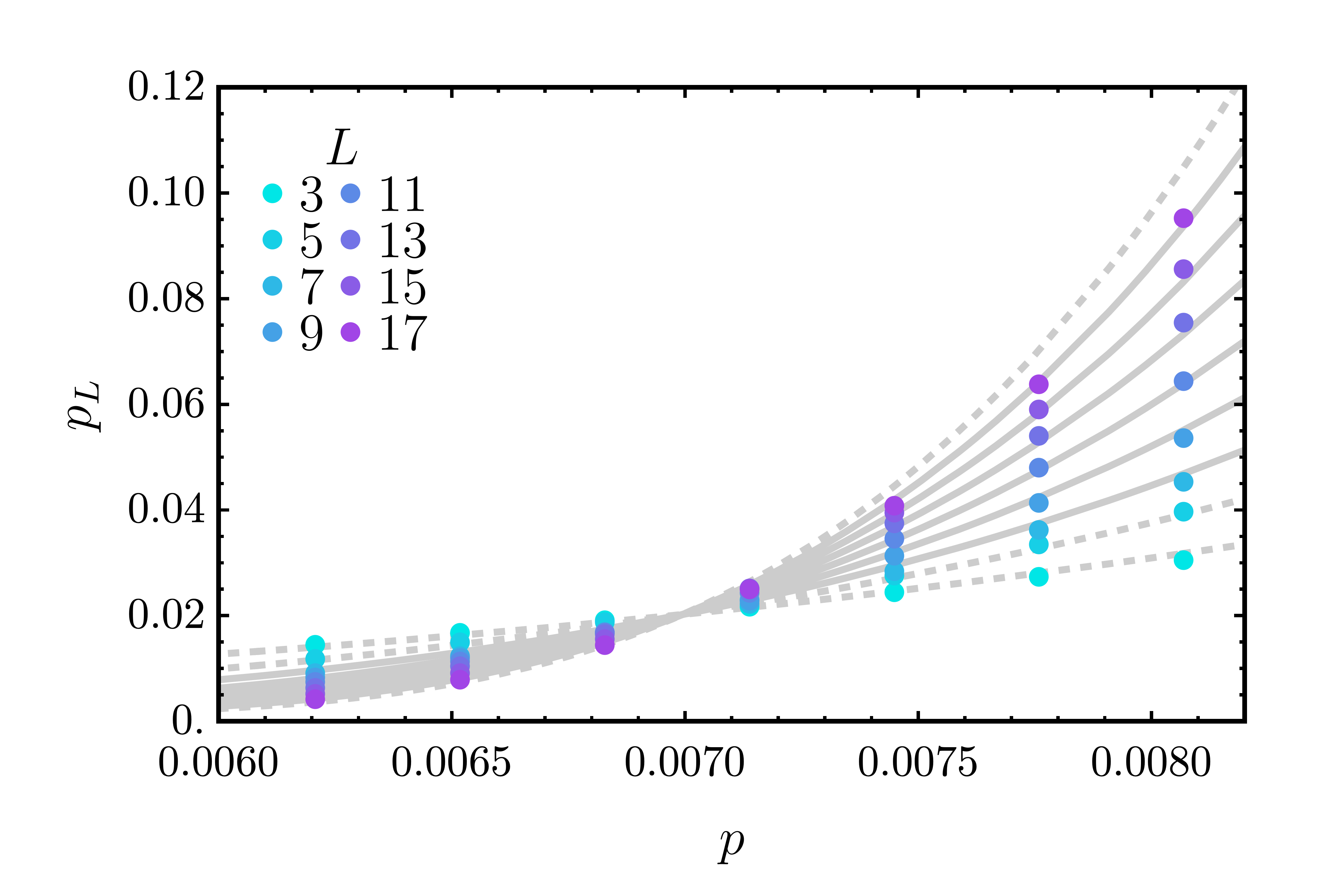}
        \caption{Logical error probability of the $\llbracket4,1,1,2\rrbracket$ scheme under the circuit-level noise model.}
    \end{minipage}%
    \label{fig:example2}%
    
\end{figure}%

\begin{figure}[h]
\centering
    \begin{minipage}[t]{0.45\textwidth}
        \includegraphics[width=\linewidth]{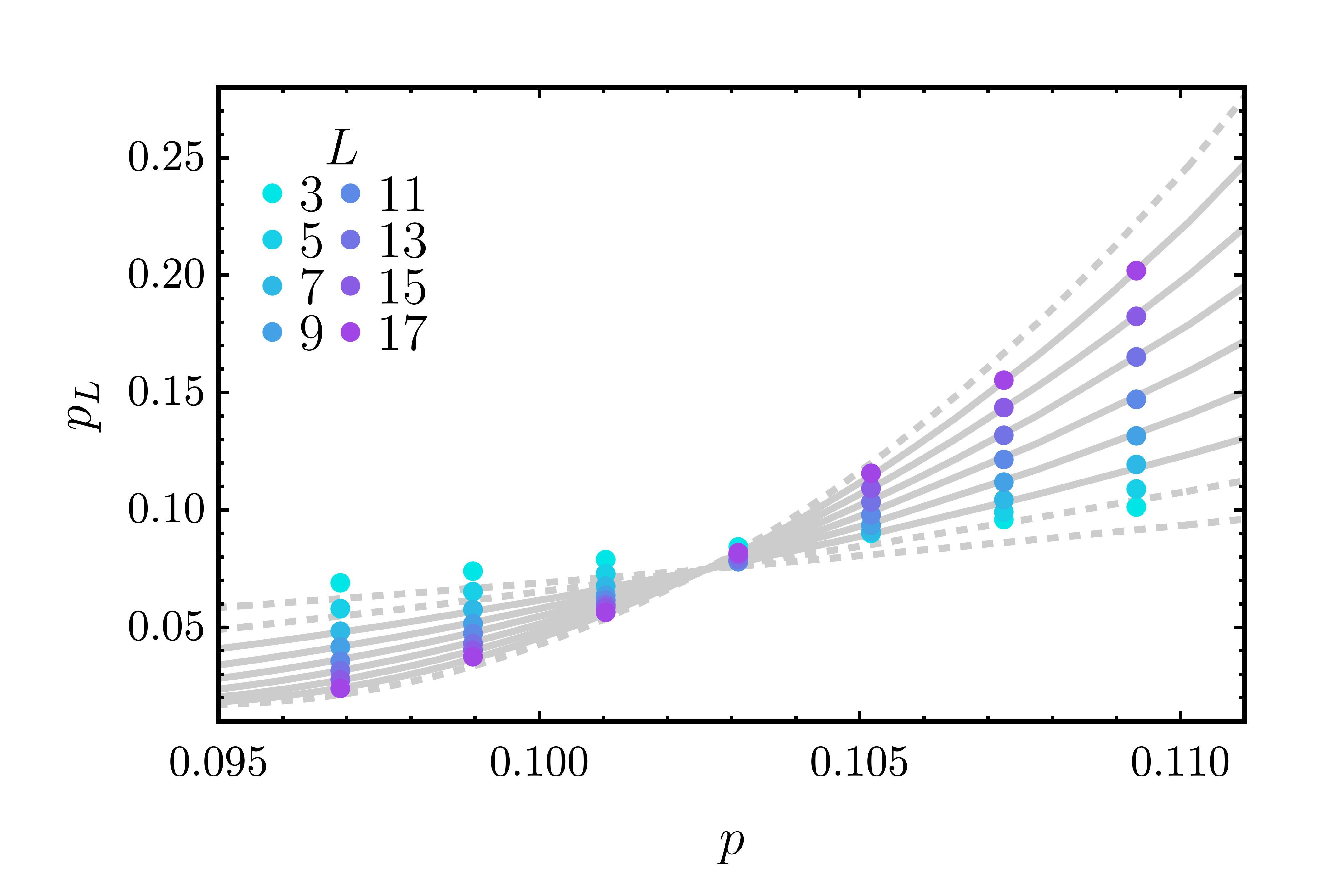}
        \caption{Logical error probability of the $\llbracket3,1,1\rrbracket_1$ scheme under the phenomenological noise model.}
    \end{minipage}\quad
    \begin{minipage}[t]{0.45\textwidth}
        \includegraphics[width=\linewidth]{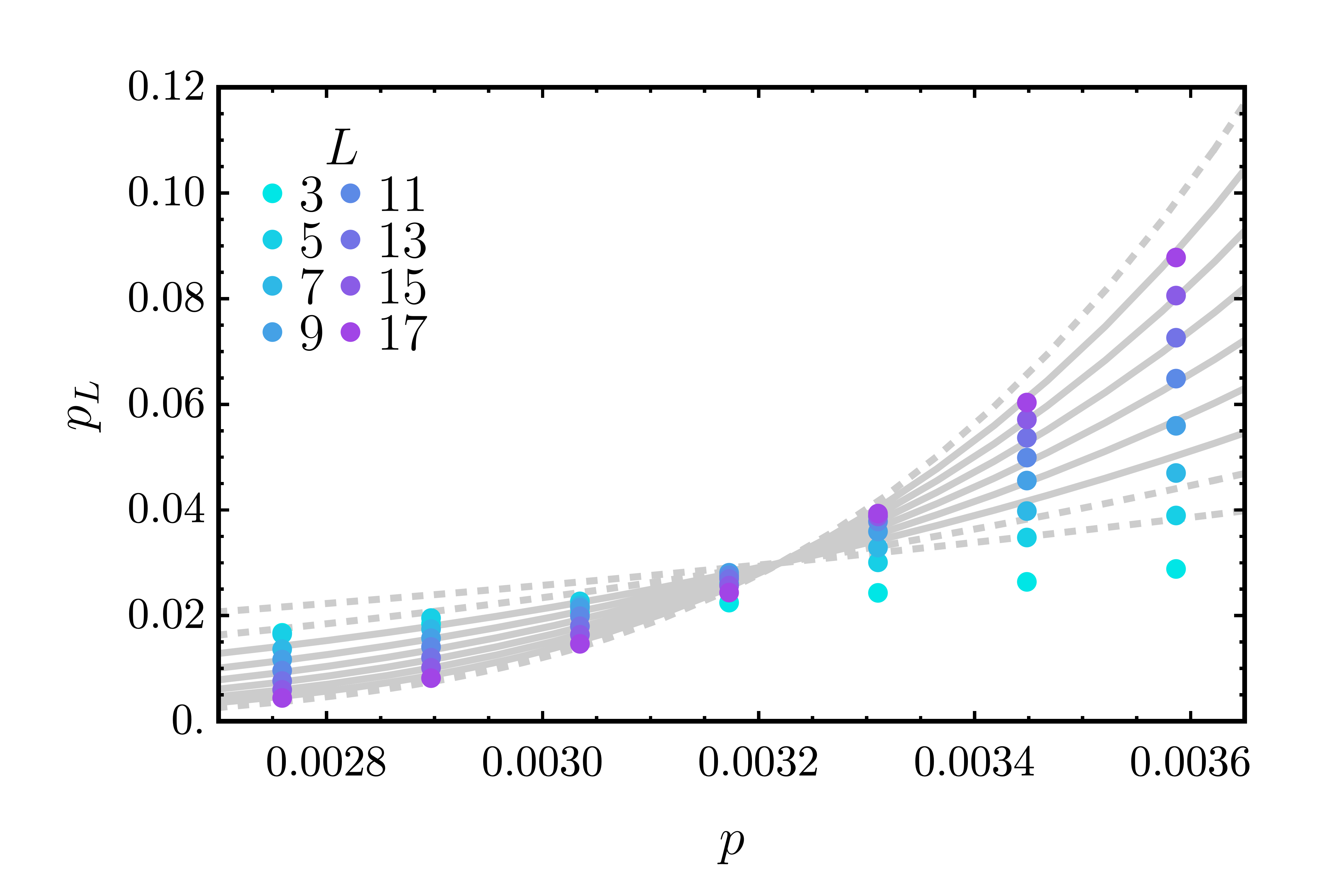}
        \caption{Logical error probability of the $\llbracket3,1,1\rrbracket_1$ scheme under the circuit-level noise model.}
    \end{minipage}%
    \label{fig:example4}%
\end{figure}

\begin{figure}[h]
\centering
    \begin{minipage}[t]{0.45\textwidth}
        \includegraphics[width=\linewidth]{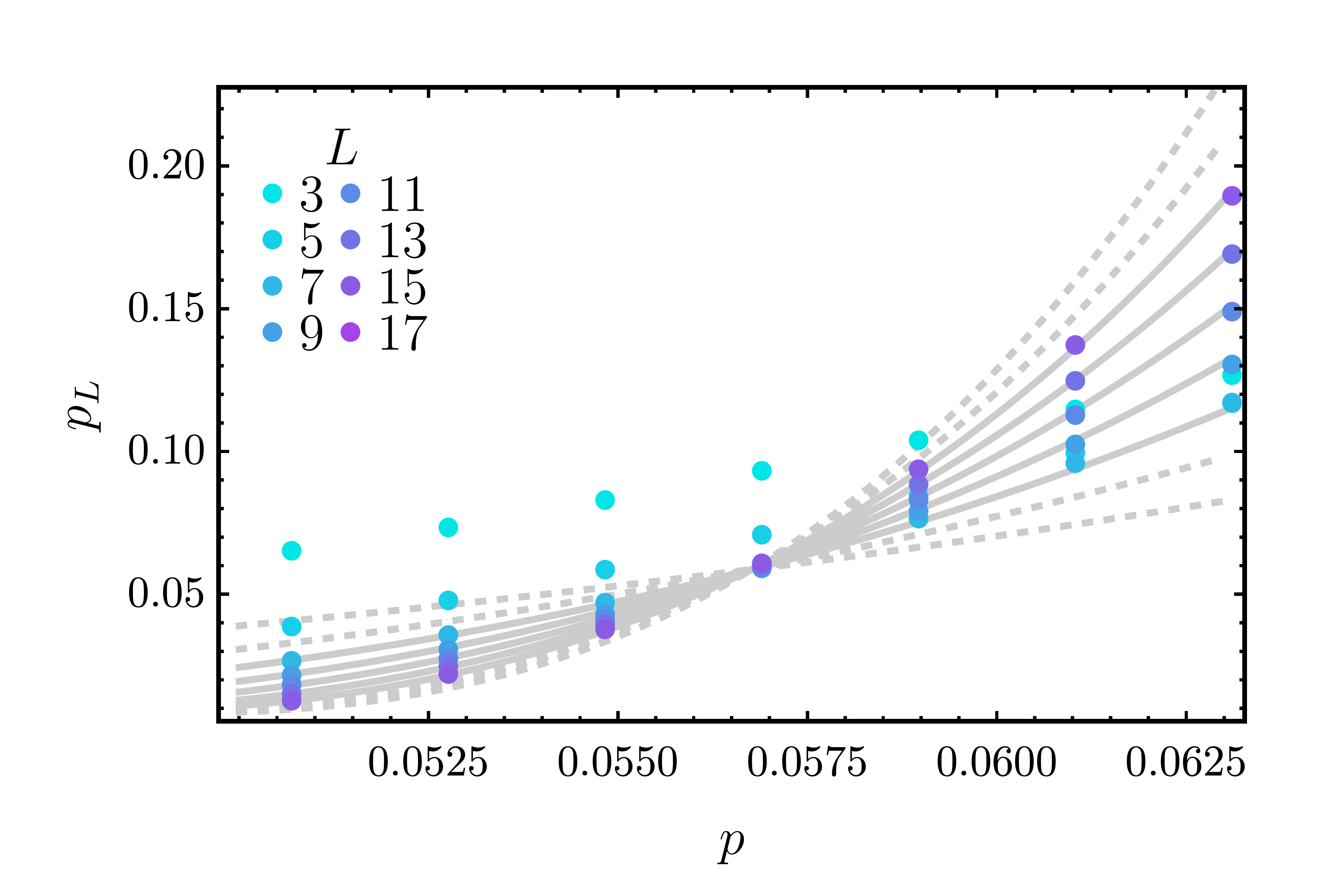}
        \caption{Logical error probability of the $\llbracket3,1,1\rrbracket_2$ scheme under the phenomenological noise model.}
    \end{minipage}\quad
    \begin{minipage}[t]{0.45\textwidth}
        \includegraphics[width=\linewidth]{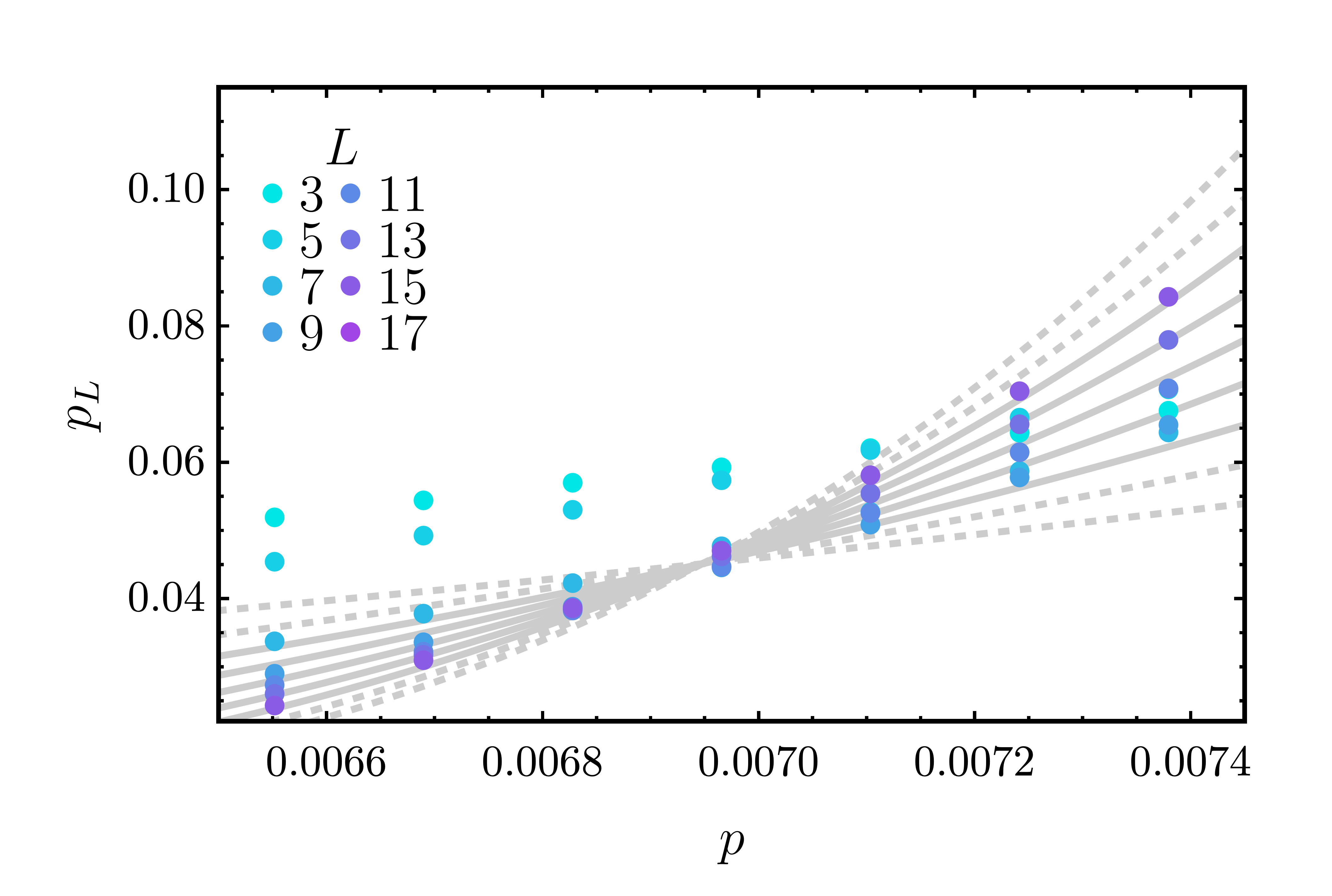}
        \caption{Logical error probability of the $\llbracket3,1,1\rrbracket_2$ scheme under the circuit-level noise model.}
    \end{minipage}%
    \label{fig:example5}%
\end{figure}\quad

\begin{figure}[h]
\centering
    \begin{minipage}[t]{0.45\textwidth}
        \includegraphics[width=\linewidth]{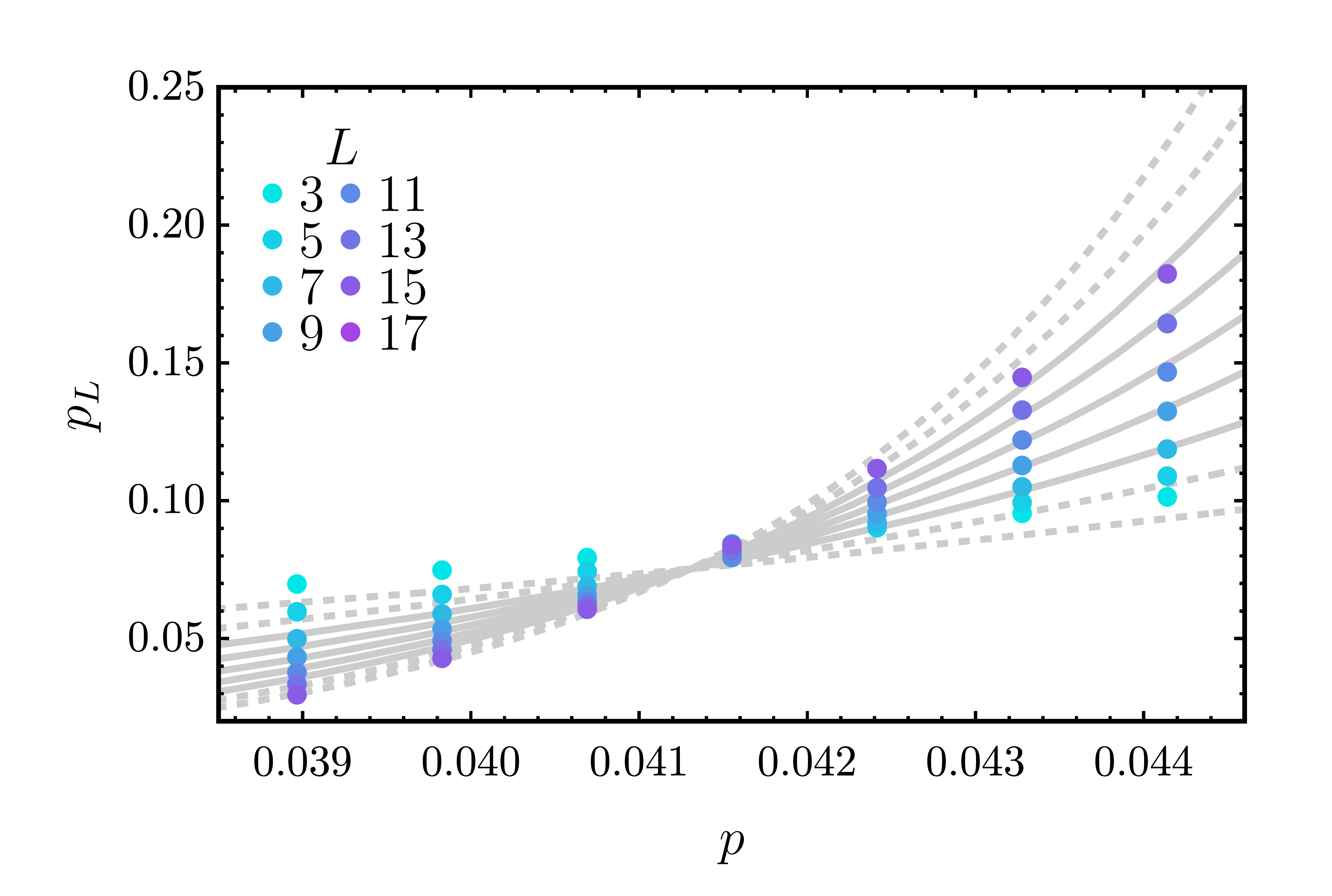}
        \caption{Logical error probability of the $\llbracket7,1,3\rrbracket$ scheme under the phenomenological noise model.}
    \end{minipage}\quad
    \begin{minipage}[t]{0.45\textwidth}
        \includegraphics[width=\linewidth]{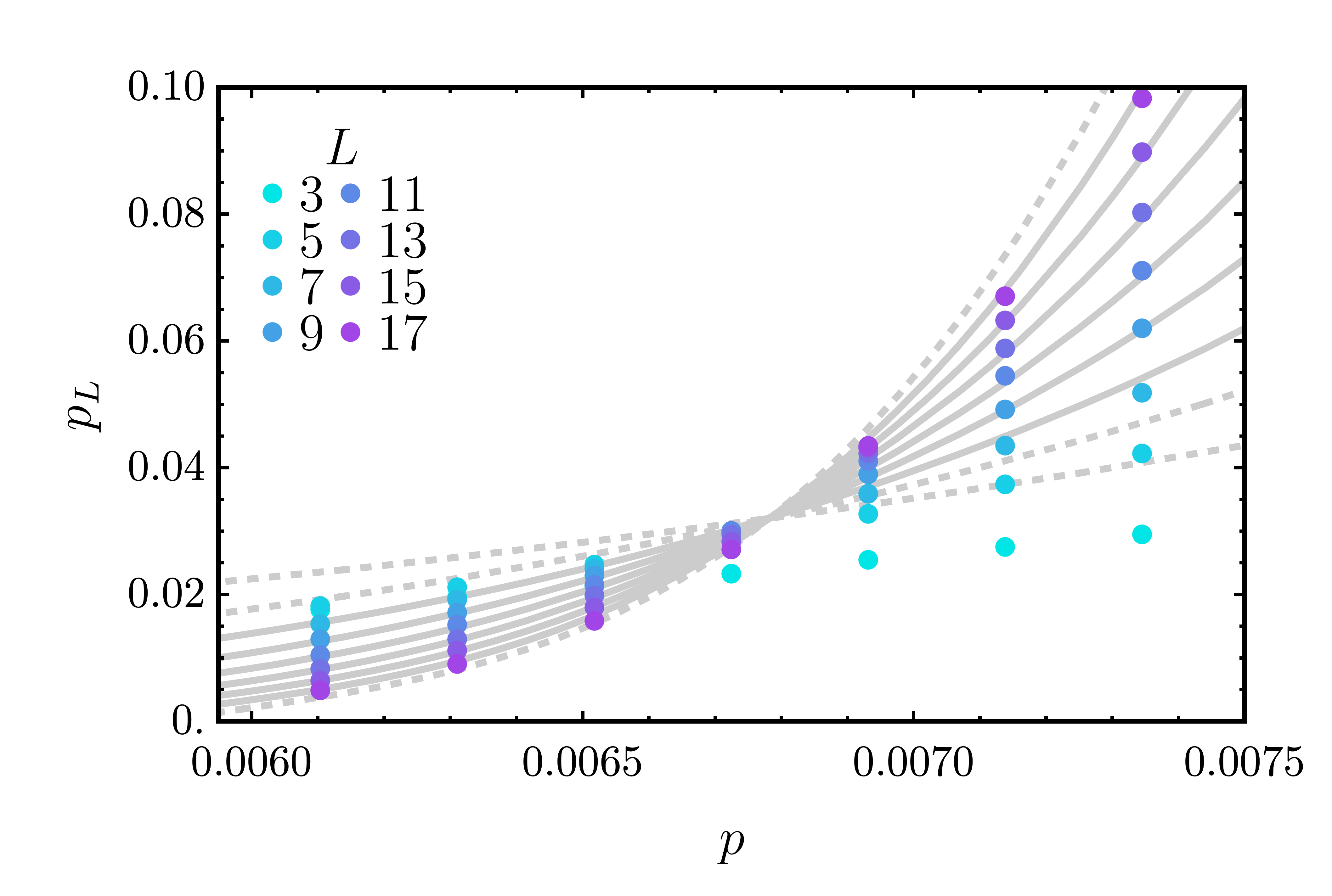}
        \caption{Logical error probability of the $\llbracket7,1,3\rrbracket$ scheme under the circuit-level noise model.}
    \end{minipage}%
    \label{fig:example6}%
\end{figure}

\twocolumngrid
\end{spacing}

\end{document}